\documentclass[iop]{emulateapj}

\usepackage{aas_macros}

\usepackage[colorlinks,citecolor=blue]{hyperref}

\usepackage{amssymb}

\usepackage{enumitem}

\usepackage[cm]{sfmath}

\usepackage{multirow}

\shorttitle{Energy budgets of giant impacts}
\shortauthors{Carter et al.}

\begin{document}

\title{The energy budgets of giant impacts}

\author{P. J. Carter$^{1}$, S. J.  Lock$^{2, 3}$ and S. T. Stewart$^{1}$}
\affil{$^{1}$Department of Earth and Planetary Sciences, University of California Davis, One Shields Avenue, Davis, CA 95616, USA\\
   $^{2}$Department of Earth and Planetary Sciences, Harvard University, Cambridge, MA 02138, USA\\
   $^{3}$Division of Geological and Planetary Sciences, California Institute of Technology, Pasadena, CA 91125, USA\vspace{1mm}\\
   {\it Accepted for publication in JGR: Planets}}
\email{pjcarter@ucdavis.edu}

\begin{abstract}
Giant impacts dominate the final stages of terrestrial planet formation and set the configuration and compositions of the final system of planets. A giant impact is believed to be responsible for the formation of Earth's Moon, but the specific impact parameters are under debate. Because the canonical Moon-forming impact is the most intensely studied scenario, it is often considered the {archetypal} giant impact. However, a wide range of impacts with different outcomes are possible. Here we examine the total energy budgets of giant impacts that form Earth-mass bodies and find that they differ substantially across the wide range of possible Moon-forming events. We show that gravitational potential energy exchange is important, and we determine the regime in which potential energy has a significant effect on the collision outcome. Energy is deposited heterogeneously within the colliding planets, increasing their internal energies, and portions of each body attain sufficient entropy for vaporization. After gravitational re-equilibration, post-impact bodies are strongly thermally stratified, with varying amounts of vaporized and supercritical mantle. The canonical Moon-forming impact is a relatively low energy event and should not be considered {the archetype} of {accretionary} giant impacts that form Earth-mass planets. After a giant impact, bodies are significantly inflated in size compared to condensed planets of the same mass, and there are substantial differences in the magnitudes of their potential, kinetic and internal energy components. 
{As a result, the conditions for metal-silicate equilibration and the subsequent evolution of the planet may vary widely between different impact scenarios.}\\

{\small KEY POINTS:}
\begin{itemize}[itemsep=0mm,leftmargin=17mm,rightmargin=15mm]
\vspace{-2mm}
\item Giant impacts involve huge exchanges between gravitational potential energy, kinetic energy and internal energy.
\item Heterogeneous internal energy increases cause vaporization of portions of the mantles of the colliding bodies.
\item Post-impact bodies are substantially inflated, with large variations in their energy components between different impacts.
\item The canonical Moon-forming impact {should not be viewed as the archetypal} giant impact.\\
\vspace{-3mm}

\end{itemize}

{\small PLAIN LANGUAGE SUMMARY:}\\
{Collisions between large planetary bodies, known as giant impacts,} dominate the final stages of the formation of rocky planets like the Earth and set the configuration and compositions of the final planets. A giant impact is believed to have formed Earth's Moon, but the specific configuration of this impact is under debate. Understanding giant impacts is crucial for understanding the formation and evolution of the Earth and the Moon as well as rocky planets {around other stars}. The traditional Moon-forming impact model is often considered the archetype of a giant impact, however, a wide range of impacts with substantially different outcomes are possible. In this work, we examine the total energies involved in giant impacts that form Earth-like planets and find that there are large differences across the wide range of possible impacts. {The internal energy increases cause large portions of each body to vaporize} as the result of impacts. Giant impacts produce planetary bodies that are significantly inflated in size compared to condensed planets of the same mass, and there are substantial differences in their potential, kinetic and internal energies. {As a result, how planets and their cores evolve after different impact scenarios may vary widely.}\\

\end{abstract}



\section{Introduction}

The final stages of terrestrial planet formation are thought to be dominated by a series of energetic impacts. Through these giant impacts, a set of lunar to Mars mass planetary embryos assemble the final system of planets \citep[e.g.][]{Chambers98}, while continuing to accrete a small percentage of their masses from leftover planetesimals and impact debris. Giant impacts sculpt planetary systems and dictate the physical and geochemical properties of the final planets. For example, in our own solar system, a giant impact with the proto-Earth is believed to be responsible for the formation of the Moon \citep[e.g.][]{Hartmann+Davis,Cameron+Ward}; and giant impacts may be responsible for {the formation of the Martian moons \citep[e.g.][]{Rosenblatt16},} Mercury's large core \citep[e.g.][]{Benz07}{, Mars' crustal dichotomy \citep[e.g.][]{Marinova08},}  {the Pluto-Charon system \citep[e.g.][]{Canup05}}, and the obliquity of Uranus \citep{Korycansky90}.

Giant impacts are high energy events. 
The outcomes of giant impacts are sensitive to the impact parameters, and impacts can have a variety of different outcomes, from merging to hit-and-run \citep[e.g.][]{Asphaug06,Genda12,Emsenhuber19}, and even erosion \citep{Leinhardt12,Stewart12}.  
Giant impacts produce a range of post-impact bodies, which can be substantially vaporized and/or rapidly rotating \citep{Lock17}. 
The evolution of the energy budget and the relative magnitude of its components during giant impacts differs greatly across the range of possible impacts.  The amount of energy that giant impacts deposit into growing bodies, and where this energy is deposited, has important consequences for the thermal and dynamical states of young planets and is a key factor in determining the composition of {any satellites that are formed} \citep[e.g.][]{Nakajima15,Lock18, Lock19}. 
It is therefore important to examine how the energy budgets of different impacts affect growing planets.

Due to the wealth of geochemical data, the Moon-forming impact has been the subject of many studies \citep[e.g.][]{Canup00,Asphaug14,Barr16} and is often considered the archetypal giant impact. In the \textit{canonical} model \citep{Canup01,Canup04,Canup08}, a Mars-sized projectile collides with the proto-Earth close to escape velocity, launching a disk into orbit around the Earth, from which the Moon accretes. This scenario is fine-tuned in order to produce a massive disk with a low iron fraction, but has difficulty explaining the isotopic similarity of the Earth and the Moon \citep[e.g.][]{Dauphas14,Dauphas17}. {Recently, \citet{Hosono19} suggested an impact onto a proto-Earth with a magma ocean would enhance the mixing and overcome the isotopic problems; however, their work also suggests that it is difficult to make a disk with sufficient mass to make a moon in the canonical scenario. Due to the difficulty of reproducing many components of the Earth-Moon system in the canonical model \citep{Asphaug14,Barr16}, several alternative models have been developed.} {One model proposes that the Moon formed via a series of lower energy impacts that produce small moonlets, which subsequently merge \citep[e.g.][]{Rufu17}.} The \textit{similar mass impactor} scenario is characterized by two approximately half-Earth-sized impactors colliding in a graze-and-merge event \citep{Canup12}. In the \textit{pre-spinning proto-Earth} model \citep{Cuk12}, a half-Mars to Mars-sized projectile hits a rapidly-rotating Earth-mass body.  The two alternative {single-}impact scenarios involve higher angular momenta than the canonical model, and lead to a greater degree of mixing between Theia (the impactor) and the proto-Earth. These alternative impacts are also fine-tuned, in this case to give a high degree of mixing, but they are just two examples of a wide range of possible Moon-forming impacts in which the Moon condenses from a cooling synestia \citep{Lock17,Hollyday17,Lock18} -- a vaporized, extended structure -- rather than forming in a classical thin, liquid-dominated disk.

Here we explore the energy budgets of accretionary giant impacts {that form Earth-mass bodies} in detail, examining the changes in internal, kinetic and gravitational potential energy as the impacts proceed. We have simulated a wide range of giant impacts that produce approximately Earth-mass final bodies, varying the mass ratios, impact angles, and impact velocities \citep{Lock17}. These collisions cover the range of terminal giant impacts onto large planetary embryos found in an example set of $N$-body simulations \citep{Quintana16} and contains examples of possible Moon-forming events including: canonical \citep{Canup01}, similar mass impactors \citep{Canup12}, and rapidly spinning proto-Earth models \citep{Cuk12}. {Hit-and-run impacts are substantially different from accretion/disruption regime impacts and are not considered here, though it should be noted that they have also been suggested as possible Moon-forming impacts \citep{Reufer12}.} We begin by discussing the general outcomes of giant impacts, and then examine the total energy budget of these events during the impact and the subsequent equilibration of the post-impact bodies. We then derive the regime in which gravitational potential energy becomes a significant factor in the energy budget. We end with a discussion of the heating of the mantle and core during giant impacts.


\section{Numerical Methods}

We examine impact simulations that were carried out using a modified version of the smoothed particle hydrodynamics (SPH) code GADGET-2 \citep{Lock17,Cuk12,Marcus09}. This code uses tabulated equations of state (EOS) for iron and forsterite ({ANEOS/}MANEOS, \citealt{Melosh07,Cuk12}) to model the cores and silicate portions of strengthless planetary embryos. {These simulations were carried out as part of several previous works \citep{Lock17,Lock19a}.} A summary of these simulations is tabulated in the supplementary materials {(Table S1)}. The modified GADGET-2 code and EOS tables are available in the online supplement of \citet{Cuk12}. {For many of the simulations in this work, a more finely gridded forsterite EOS table was used, this table is available from \citet{Carter19GIDATA}.}

The targets range in mass from 0.52--1.05\,M$_\oplus$, and have a minimum resolution of 100,000 particles, and the projectiles range from 0.03--0.52\,M$_\oplus$, with a minimum of 5000 particles of the same mass as those in the corresponding target. All bodies have an iron core and a forsterite mantle, with a core mass fraction of 0.3. The sizes of targets and impactors, impact velocities and impact parameters were chosen to sample both the range of impact scenarios proposed for the Moon-forming giant impact, and the distribution of impact parameters for terminal impacts {with projectiles at least as massive as 1\% of the mass of the Earth} from an example set of $N$-body simulations of the late stages of terrestrial planet formation \citep[e.g.][figure 13]{Quintana16}. The bodies begin sufficiently close together that there is minimal deformation and the velocity does not change significantly prior to the impact.

\begin{figure*}
    \centering
    \includegraphics[width=1.\textwidth]{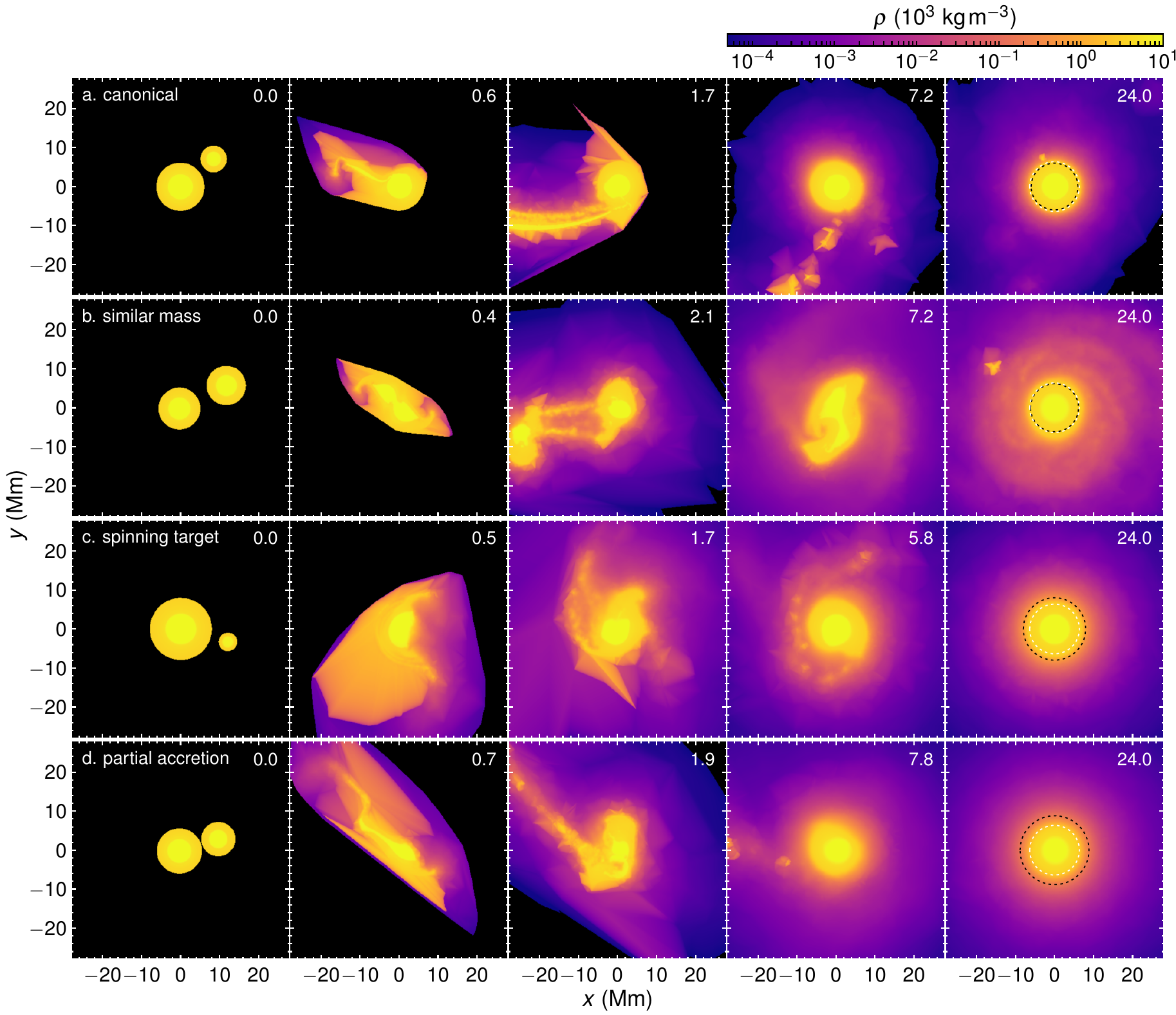}
    \caption{Time sequences of impacts showing density of the SPH simulations in the equatorial plane. (a) A canonical Moon-forming impact between a 0.9\,M$_\oplus$ target and a 0.13\,M$_\oplus$ projectile {(M0.9L0T2000M0.13L0T2000v9.2b0.74)}; (b) a similar mass impactors Moon-forming scenario involving {bodies} with masses of 0.57\,M$_\oplus$ and 0.47\,M$_\oplus$ {(M0572p0M468p0v9.7b0.55)}; (c) a pre-spinning proto-Earth Moon-forming impact between a 1.05\,M$_\oplus$ target with an angular momentum of 2.7 L$_{\rm EM}$ (spin period of 2.4\,hours{;  L$_{\rm EM}$ is the present-day angular momentum of the Earth-Moon system}) and a 0.05\,M$_\oplus$ projectile {(C105p2.4M0.05v20br0.3)}; and (d) a partial accretion impact between a 0.75\,M$_\oplus$ target and a 0.3\,M$_\oplus$ projectile {(M0.75m0.3v1.25b0.3)}.  The time in hours is shown in the top right corner of each panel. The density of the fluid is shown using a Delaunay triangulation interpolation (which results in edge effects in regions with few/no particles), rather than showing the individual SPH particles which do not represent the low density regions well for these continuum calculations. Images are recentered on the gravitational potential minimum of the system in each panel. Black regions are below the minimum density on the color scale. The dashed white circles {in the final column (obscured by the black circles in the top two rows)} indicate the present-day size of the Earth, with radius 6.37\,Mm; the black circles show the size of the corotating regions at the end of these simulations [these have radii: a) 6.0\,Mm, b) 6.2\,Mm, c) 8.0\,Mm, and d) 8.8\,Mm]. Accompanying animations are provided in the supplementary materials.}
    \label{f:impactseq}
\end{figure*}

We use runs 118 {(M0.9L0T2000M0.13L0T2000v9.2b0.74)}, 129 {(M0572p0M468p0v9.7b0.55)}, 1 {(C105p2.4M0.05v20br0.3)} and 159 {(M0.75m0.3v1.25b0.3)} from table S4 of \citet{Lock17} throughout this article as examples of possible Moon-forming impacts: the canonical, similar mass impactors, pre-spinning proto-Earth scenarios; and a \textit{partial accretion} scenario representing {an} example of the {high energy} accretionary giant impacts expected {between large embryos} during the final stages of {terrestrial} planet formation. Time sequences for these example impact simulations are shown in Figure \ref{f:impactseq}. The details of the example simulations are provided in Table \ref{t:simtable}, and details of the entire set (which includes more examples of terminal impacts based on \citealt{Quintana16} than were included in \citealt{Lock17} and \citealt{Lock19a}) are provided in the supplementary materials {(Table S1)}.

{In this work, we found an increase in the total energy budget of the SPH simulations, generally on the order of 2 to 5 percent. We report the errors as a fraction of the energy budget using the participating potential energy (defined below) as this provides a better reference for energy error than absolute total energy, which can be very close to zero due to the negative potential energy. The SPH simulations used ANEOS/MANEOS material models that were tabulated on a density-entropy grid. The accuracy of the interpolation is variable across the grid, and we found that the largest errors are associated with interpolations within the vapor dome. In the impact events with the largest specific energies, the vaporizing ejecta traverses the vapor dome region of the EOS table and accumulates error in the energy budget. We found that the energy conservation of the bound mass, calculated for times after the initial contact and deformation stage, was much better.}

{Giant impacts that form an Earth-mass body have high specific energies and most of the ejected material is substantially vaporized. We found that, in the majority of the simulations, the energy increases by less than 5\% of the total energy. For the four example impacts described above, the total energy budget increases by between 3 and 16\%. We recalculated the example case with the worst energy conservation (similar mass) and the canonical case with a finer resolution grid in the vapor dome region and the errors reduced to 1 and 0.5\% respectively (see Supplementary Figure S2).  
We also examined calculations of isolated synestias where all the mass is bound (from \citet{Lock17} using the same EOS tables as in this work). As the synestias viscously spread, the total energy budgets were conserved to within 0.5\%.}

{In this work, we focus on the energy budget during the event and the exchange of energy between different components in the bound mass. The error that accumulates within the escaping ejecta, while unfortunate, does not affect the processes discussed below and the magnitude of the error is much smaller than the gain in internal energy in the final body. Future work will use improved gridding and interpolation schemes to minimize the errors associated with using tabulated EOS.}

The properties of comparison planets at the magma-ocean stage were calculated using the HERCULES code \citep{Lock17,HERCULES} with the same EOS as used in the SPH simulations. {HERCULES uses a potential field method to calculate the equilibrium structure of a body with a given thermal state, composition, mass and angular momentum.} The magma-ocean planets were assumed to have isentropic cores and mantles with specific entropies of 1.5 and 4\,kJ\,K$^{-1}$\,kg$^{-1}$ respectively. This core isentrope has a temperature of 3800\,K at the pressure of the present-day core-mantle boundary, similar to the present thermal state of Earth's core. The mantle isentrope intersects the liquid-vapor phase boundary at low pressure {(1\,MPa)} and about 4000\,K. Our chosen thermal state approximates that of a well-mixed, mostly-liquid, magma-ocean planet. For this paper, we used the same HERCULES parameters as used in \citet{Lock17}.

\begin{turnpage}
\begin{table*}
\footnotesize
\caption{Summary of example simulations.\\ $M$ and $R$ are the mass and radius of the target, $L$ and $\omega$ are the angular momentum and angular velocity of the target, $m$ and $r$ are the mass and radius of the projectile, $v_\mathrm{i}$ is the impact velocity, $v_\mathrm{esc}$ is the mutual escape velocity, $b$ is the impact parameter, $Q_{\rm S}$ is the specific impact energy (see \ref{a:qs}), $M_{\rm bnd}$ and $L_{\rm bnd}$ are the mass and angular momentum of the post-impact body, $E_\mathrm{pot, 0}$ and $E_\mathrm{pot, min}$ are the initial and minimum potential energy and $\Delta E_\mathrm{int}$ is the gain in internal energy at 24 hours. L$_{\rm EM}$ is the present-day angular momentum of the Earth-Moon system. Full details for all simulations in the impact database are given in the supplementary material (Table S1). \label{t:simtable}}
\centering
\begin{tabular}{lccccccccccccccc}
\hline
Impact & $M$ & $R$ & $L$ & $\omega$  & $m$ & $r$ & $v_\mathrm{i}$ & \multirow{2}{*}{$\displaystyle \frac{v_\mathrm{i}}{v_\mathrm{esc}}$} & $b$  & $Q_{\rm S}$ & $M_{\rm bnd}$ & $L_{\rm bnd}$ & $E_\mathrm{pot, 0}$ & $E_\mathrm{pot, min}$ & $\Delta E_\mathrm{int}$\\
 & (M$_\oplus$) & (km) & (L$_{\rm EM}$) & (10$^{-3}$ rad\,s$^{-1}$)   & (M$_\oplus$) & (km) & (km$\,$s$^{-1}$)  & & & (J\,kg$^{-1}$) &  (M$_\oplus$) & (L$_{\rm EM}$) & (10$^{32}$ J) & (10$^{32}$ J) & (10$^{32}$ J)\\
\hline
Canonical{\footnote{{M0.9L0T2000M0.13L0T2000v9.2b0.74}}}        & 0.90  & 6160 & 0 & 0   & 0.13  & 3390 & 9.2  & 0.99 & 0.74     & 4.61$\times 10^5$ & 0.99 & 0.85 & -2.391 & -2.559 & 0.226 \\
Similar-mass{\footnote{{M0572p0M468p0v9.7b0.55}}}     & 0.57  & 5350 & 0 & 0   & 0.47  & 5040 & 9.7  & 1.09 & 0.55     & 5.66$\times 10^6$ & 0.97 & 2.16 & -2.087 & -2.589 & 0.534 \\
Spinning proto-Earth{\footnote{{C105p2.4M0.05v20br0.3}}}  & 1.05 & 7980 & 2.69 & 0.72   & 0.05  & 2380 & 20  & 2.18  & $-$0.3   & 6.42$\times 10^6$ & 1.03 & 2.10 & -2.686 & -2.802 & 0.394 \\
Partial accretion{\footnote{{M0.75m0.3v1.25b0.3}}}  & 0.75  & 5800 & 0 & 0   & 0.30  & 4400 & 11.3 & 1.25 & 0.3      & 1.20$\times 10^7$ & 1.02 & 1.14 & -2.337 & -2.751 & 0.633 \\
\hline
\end{tabular}
\end{table*}
\end{turnpage}
%


\section{Collision outcomes}

The impacts in our database span a range of collision outcomes from partial accretion to erosion and include graze-and-merge events. The majority of the impacts lead either to accretion of a large fraction of the impactor mass or slight erosion of the target. It should be noted that in this context the material that eventually forms the Moon is bound to the post-impact Earth, and is therefore considered to be accreted. 

\begin{figure}
    \centering
    \includegraphics[width=\columnwidth]{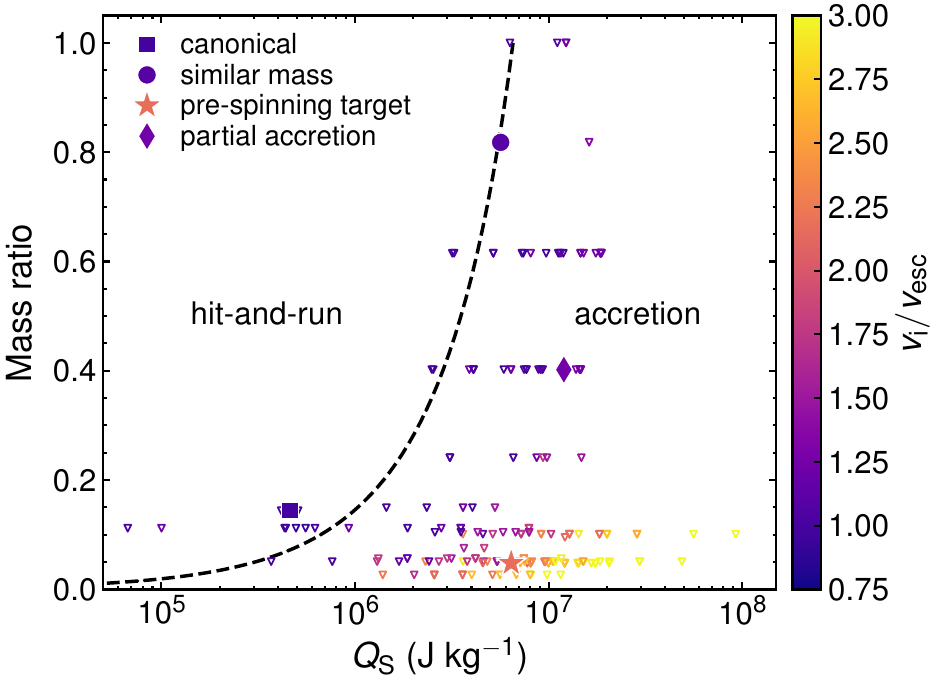}
    \caption{Projectile-to-target mass ratio vs specific impact energy for the impacts in our database. The dashed line indicates the transition between non-grazing and grazing impacts for impacts at the mutual escape velocity between bodies with a combined mass equal to the mass of the Earth {\citep[based on][]{Leinhardt12}}. The impacts just to the left of this line are in the graze-and-merge regime. The colors of the points indicate the impact velocity. The four example impacts shown throughout this work are indicated by unique symbols.}
    \label{f:grazetransition}
\end{figure}

The $N$-body simulations conducted by \citet{Quintana16} show that the high energies of giant impacts are achieved across a large range of projectile-to-target mass ratios \citep[see figure 13 in][]{Quintana16}. The mass ratios and modified specific impact energies, $Q_\mathrm{S}$ (\citealt{Lock17}; see \ref{a:qs} for the definition), for the giant impacts in our database are shown in Figure \ref{f:grazetransition}. {These simulations represent possible single event collision outcomes that produce approximately Earth-mass final bodies, and Figure \ref{f:grazetransition} does not indicate collision probabilities. The probabilities of collision outcomes depend on the context of planet formation; a model that provides mass ratio and impact velocity distributions can be used to calculate probabilities, as was done in \citet{Stewart12}}.

The dashed line in Figure \ref{f:grazetransition} shows the transition from accretionary impacts to hit-and-run impacts for impacts at the mutual escape velocity between bodies with a combined mass equal to the mass of the Earth. {This line was obtained from the critical impact parameter as defined in \citet{Leinhardt12}}. Lower impact energies require lower impact velocities and/or larger impact angles (such that a smaller fraction of the mass of the projectile participates in the impact, see \citealt{Leinhardt12}). Just to the left of the line is the graze-and-merge regime, which includes the canonical Moon-forming impact (blue square). In general, impacts to the left of this transition line are likely to be hit-and-run impacts, and thus not affect the mass of the target body. {Figure \ref{f:grazetransition} demonstrates that giant impacts cannot be defined solely via an energy criterion.} 
Typical giant impacts {with a combined mass equal to that of the Earth} that lead to growth and/or moon formation {are found} on the right side of this line. {In addition, many of the impact events that occur in the $N$-body simulations from \citet{Quintana16} fall on the right hand side of the dashed line in Figure \ref{f:grazetransition}. As such, we cannot take the canonical model as being typical of all giant impacts that produce an Earth-like planet.}

The canonical model for the Moon-forming impact {\citep[e.g.][]{Canup01,Canup04,Canup08}} is a graze-and-merge event, in which the projectile and target separate after the initial impact with little change in target mass, but remain bound and later re-collide and merge \citep{Stewart12}.

The similar mass impactor scenario suggested by \citet{Canup12} is also a graze-and-merge type impact, however, it is considerably different than the canonical model due to the similar mass of the two bodies, the order of magnitude greater specific energy, the greater degree of mixing{, and the larger total angular momentum}. The {colliding bodies} initially separate but remain bound and so undergo a series of impacts with reducing separation as they spiral around each other until the bodies eventually merge.

The rapidly-spinning proto-Earth model {\citep[based on][]{Cuk12}} is an example of an erosive impact: the final bound mass of the largest remnant body is slightly less than that of the original target. The high velocity projectile largely merges with the target, but the energy of this collision causes ejection of some mass from the target, as well as substantially inflating the body due to vaporization.

The final example is a {high energy} partial accretion impact. It is close to a graze-and-merge impact, but the cores do not fully separate after the initial impact. In this case approximately 10\% of the mass of the projectile escapes, and the rest is accreted. As is generally the case in giant impacts, both target and projectile are significantly deformed during the impact.

These four cases, shown in Figure \ref{f:impactseq}, serve to illustrate the diversity of possible {non-hit-and-run} giant impacts that form an approximately Earth-mass body. Most of the giant impacts in {this work} result in post-impact bodies that exceed the co-rotation limit (CoRoL) -- the thermal limit beyond which a body cannot be in hydrostatic equilibrium and have a constant angular velocity \citep{Lock17}. 
Planetary bodies that exceed the CoRoL are known as synestias. A synestia typically has a smooth transition between an inner co-rotating region and an outer disk, which has substantially sub-Keplerian angular velocities. In the example impacts discussed above, only the canonical model -- which has a much smaller specific energy and angular momentum than the other impacts (see Table \ref{t:simtable}) -- results in a sub-CoRoL post-impact body, with a strong surface density drop between the corotating `planet' and the colder near-Keplerian disk. One clear contrast between the sub-CoRoL post-impact structure and synestias is the difference in the density distribution in the post-impact bodies. This is illustrated in the final column of Figure \ref{f:impactseq}, where rows {\it b}, {\it c} and {\it d} show an extended moderate density (50--1000\,kg\,m$^{-3}$, orange) region and a low density (0.1--10\,kg\,m$^{-3}$, purple) region extending beyond the span of the panel, while row {\it a} shows a much smaller low density disk (purple with a transition to dark blue and to black where the density drops below the minimum of the color scale, 0.05\,kg\,m$^{-3}$, at the edges of the panel). For a detailed discussion of the structure of terrestrial bodies with different thermal energies and angular momenta, we refer the reader to \citet{Lock17}.


\section{Energy budgets}

The energy budget of an impact consists of three components: the gravitational potential energy, kinetic energy and internal energy. We consider the transfer of energy between these components as impacts proceed. We will first discuss the general features of the evolution of energy budgets of giant impacts and then compare our four example cases.

\subsection{Evolution of energy budgets in impacts}\label{s:budgets}

The initial energy budget of the colliding system is comprised of the gravitational potential energy, the kinetic energy of the approaching {bodies}, and the internal energies of the impacting bodies. The total initial gravitational potential energy in an impact is the sum of three components: the potential energy due to the separation of the {two bodies}, the gravitational binding energy of the target and the binding energy of the projectile. By convention the gravitational potential energy is defined as negative. {The simulations we discuss in this work begin with various small offsets between the two bodies. As such, the initial potential energy is not a good reference point, starting further away or closer to contact would change this value. Below we will adopt the minimum value of the potential energy as the useful reference value.} It is important to note that for planets with the same total mass, an inflated planet or synestia will have a {less negative} binding energy than a condensed body, as the mass is distributed at greater distances from the planet's center of mass.

\begin{figure*}
    \centering
    \includegraphics[width=0.75\textwidth]{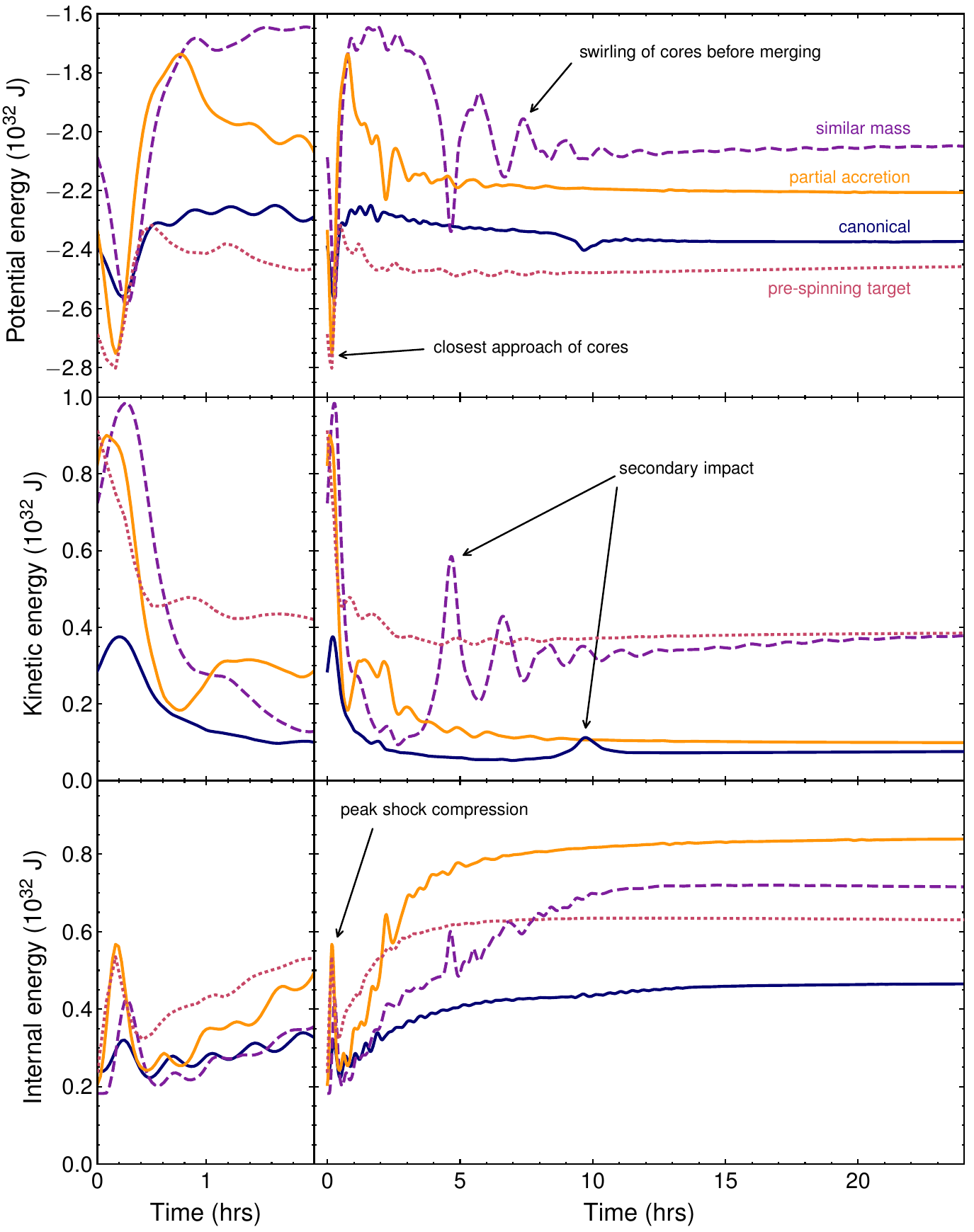}
    \caption{Energy components in the example Moon-forming impact simulations. {The left hand panels show the first 2 hours in more detail.} The canonical model is shown as a solid blue line, the similar-mass impactor scenario as a dashed purple line, the pre-spinning proto-Earth scenario with a dotted magenta line, and the partial accretion example with a solid orange line. Due to a large contribution from gravitational potential energy the total energy is negative in all the examples.}
    \label{f:rawenergy}
\end{figure*}
The total energy of the system in the Moon-forming impact scenarios {is negative due to a large contribution from potential energy (see Figure \ref{f:rawenergy}). This is the case for most non-grazing giant impacts: the majority of giant impacts lead to partial accretion in standard models of terrestrial planet formation in our solar system \citep{Stewart12}}. {In partial accretion collisions,} most of the mass remains bound and will form the post-impact body, including the material that will form the Moon. Note that material that is ejected in the impact still orbits the host star, and may re-encounter the main post-impact body on subsequent orbits {\citep[e.g.][]{Jackson12}}.

During the approach, the {bodies} fall into each other's potential wells, and the potential energy decreases (becomes more negative). The potential energy is converted into kinetic energy as the bodies accelerate towards each other (see Figure \ref{f:rawenergy}). As the impact begins, the potential energy continues to decrease as the the separation between the projectile and target mass shrinks. In some impacts, the kinetic energy also continues to increase noticeably after first contact as the {colliding bodies} accelerate towards their common center of mass. The shock caused by the collision transfers energy into the {bodies}, causing a spike in the internal energy and changing their kinetic energies.

Shortly after the impact begins, the bodies reach a state of maximum compression, when the mass is at its most concentrated. This corresponds to the minimum in potential energy. In many cases, some portion of the projectile continues moving, and travels away from the center of mass. As mass moves up the potential well, the potential energy increases again and the kinetic energy decreases due to deceleration. At the same time, the shocked material decompresses and expands, and most of the internal energy {gained} from the initial impact is converted back into kinetic and potential energy. In general, a large fraction of this expanding or separating mass is bound to the remnant, and so the system reaches a state of maximum decompression, typically about one hour after the start of the impact, after which the displaced material falls back into the potential well reducing the potential energy once again, but to a lesser degree than the initial impact. 
{As gravitational re-equilibration proceeds, expanded fluid is compressed by gravitational forces and fragments fall back onto the post-impact body generating secondary shocks. These secondary shocks cause further heating \citep{Nakajima15}. This heating due to infall leads} to a more gradual increase in internal energy. The dynamical time for these impacts is hours {($t_\mathrm{dyn} \simeq D^3 / \sqrt{G M_\mathrm{tot}}$, where $D$ is the sum of projectile and target radii, $G$ is the gravitational constant and $M_\mathrm{tot}$ is the total mass involved in the impact)}, but the time to reach gravitational equilibrium varies strongly with impact scenario.

\begin{figure*}
    \centering
    \includegraphics[width=0.96\textwidth]{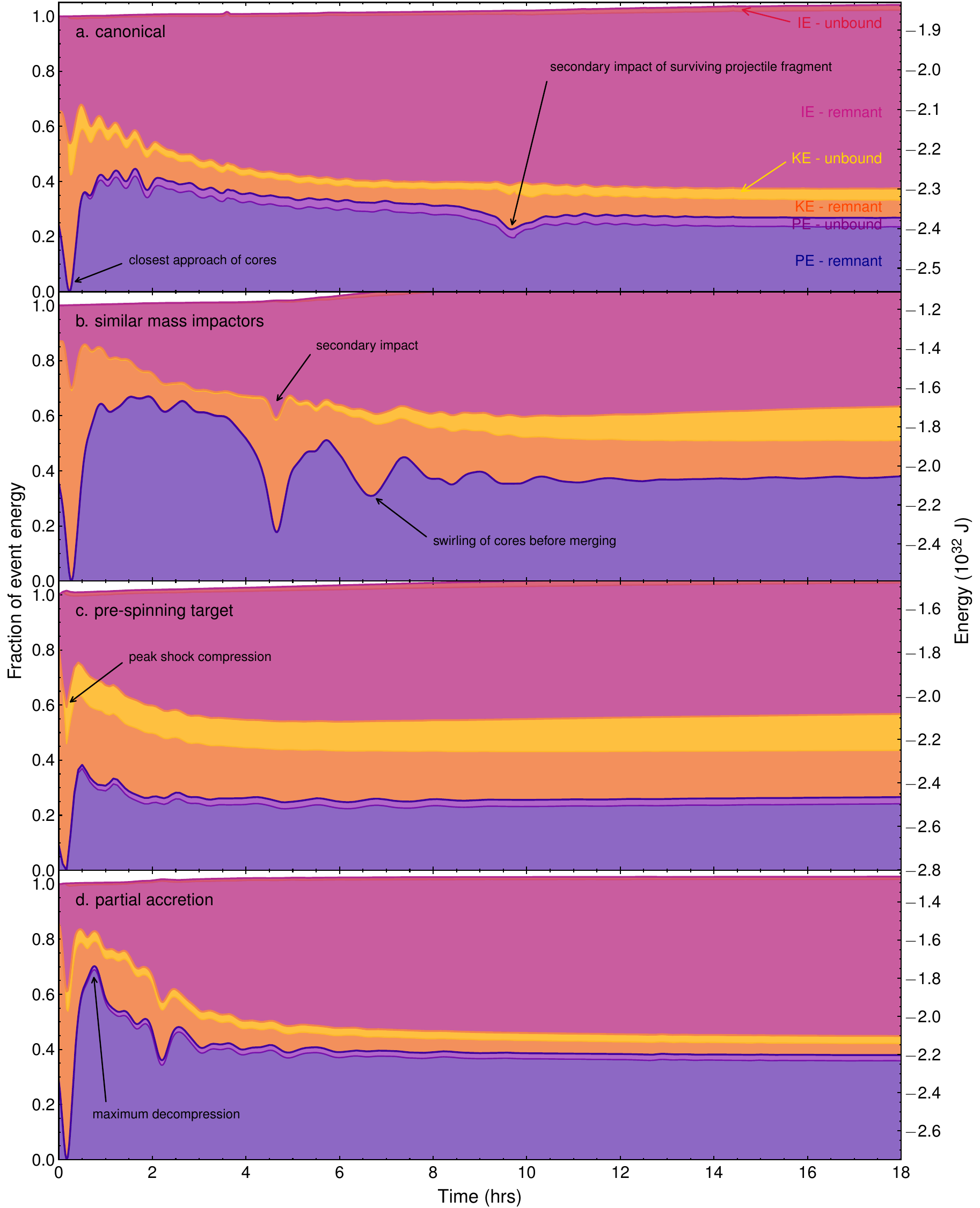}
    \caption{Evolving energy budgets of the example impacts from Figure \ref{f:impactseq}: (a) a canonical Moon-forming impact; (b) a similar mass impactors Moon-forming scenario; (c) a pre-spinning proto-Earth Moon-forming impact; and (d) a partial accretion impact. The energy budget is the sum of the 'participating' potential energy, ${E_\mathrm{pot} - E_\mathrm{pot, min}}$ (PE -- purple), kinetic energy (KE -- orange) and internal energy (IE -- magenta). The lilac (PE), yellow (KE) and pink (IE) regions indicate the portion of each energy component associated with material that is no longer bound, and in some cases are too small to see on this scale. Note that the total energy increases above the scale of the axes in panels b and c.}
    \label{f:ebudgets}
\end{figure*}

As we have seen, the energy changes are large, and there is substantial exchange between internal, kinetic, and gravitational potential energies ($E_\mathrm{pot}$) during impact events. {Potential energy is key to the dynamics of the impact as work has to be done to change the potential energy}. However, as much of the total potential energy is due to the combined potential well of the colliding bodies, it cannot participate in the impact.  For convenience, we describe the giant impact energy budget at any point in time by the kinetic energy, internal energy, and a `participating' potential energy term -- found by subtracting the minimum value of the potential energy: $E_\mathrm{pot} - E_\mathrm{pot, min}$. This offset factor $E_\mathrm{pot, min}$, which will be estimated in Section \ref{s:PEcalc}, converts the potential energy term to a positive value. Note that the potential minimum occurs during the early stages of the impact and is a more negative value than either the beginning or end states. The total energy budget is then the sum of these three terms and should be constant as the impact proceeds and the post-impact state approaches equilibrium. The energy budgets of our example impacts are shown in Figure \ref{f:ebudgets}.

If a large fragment of the projectile survives the initial impact, but is still bound, a significant secondary impact occurs (at about 9.5 hours in Figure \ref{f:ebudgets}a, and at about 4.5 hours in Figure \ref{f:ebudgets}b). In these cases this surviving portion of the projectile accelerates as it falls back down the potential well of the other body, causing a conversion of potential energy back into kinetic energy. The secondary impact often has a similar phenomenology to the initial impact, though is less energetic (the impactor has lost mass and has a lower velocity with some of the initial energy having already been converted to internal energy of the bodies). Again there is some decompression after the secondary impact. Overall secondary impacts tend to have a smaller lasting effect on the energy partitioning compared to the initial contact.

The contribution to the energy budgets from the cores of the {colliding bodies} is generally smaller than that from the mantles. As the cores are more dense, they contribute a larger fraction of the potential energy ($\sim$40\%) than the core mass fraction (30\%), however, their contribution to the internal energy is generally smaller than their mass fraction.

\subsection{Energy budgets of post-impact bodies}

\begin{figure}
    \centering
    \includegraphics[width=\columnwidth]{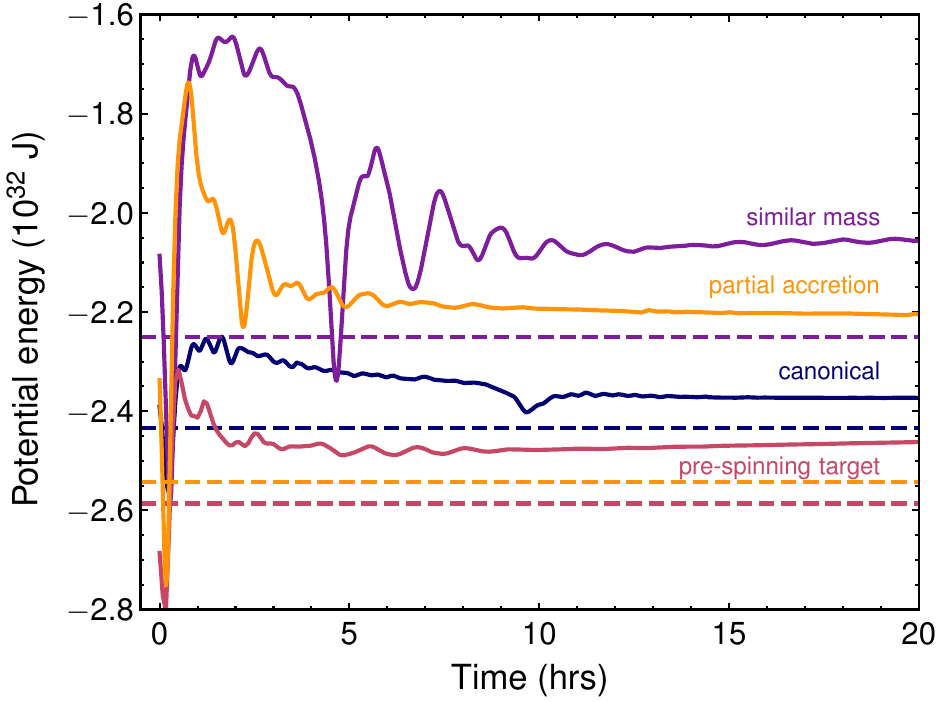}
    \caption{Potential energy evolution for example giant impacts (solid lines) compared to the potential energy of planets at the magma-ocean stage (dashed lines) with the same mass as the bound post-impact remnant calculated using the HERCULES code.}
    \label{f:potentialE}
\end{figure}
{Studies of post-impact bodies commonly assume a fully condensed magma-ocean stage planet as their starting point. However, the planetary body that exists immediately after a giant impact can be very different from a condensed planet with the same mass.} The energy budgets show that post-impact {bodies} {have a large contribution from potential energy}. We find that their gravitational potential energies after equilibration are significantly different from those of planets at the magma-ocean stage with the same masses, as shown in Figure \ref{f:potentialE}. This excess potential energy may be sufficient to melt the entire mantle of an Earth-mass planet \citep{Tonks93}. The larger potential energies of thermally inflated post-impact bodies will affect the cooling and subsequent evolution of the bodies. {The thermal structure affects the composition of the Moon formed after an impact, as discussed by \citet{Lock18}. The evolution of the energy budgets include changes in internal energy as the shape of the planet changes during cooling and tidal recession of the Moon \citep{Lock19}.}

\begin{figure}
    \centering
    \includegraphics[width=\columnwidth]{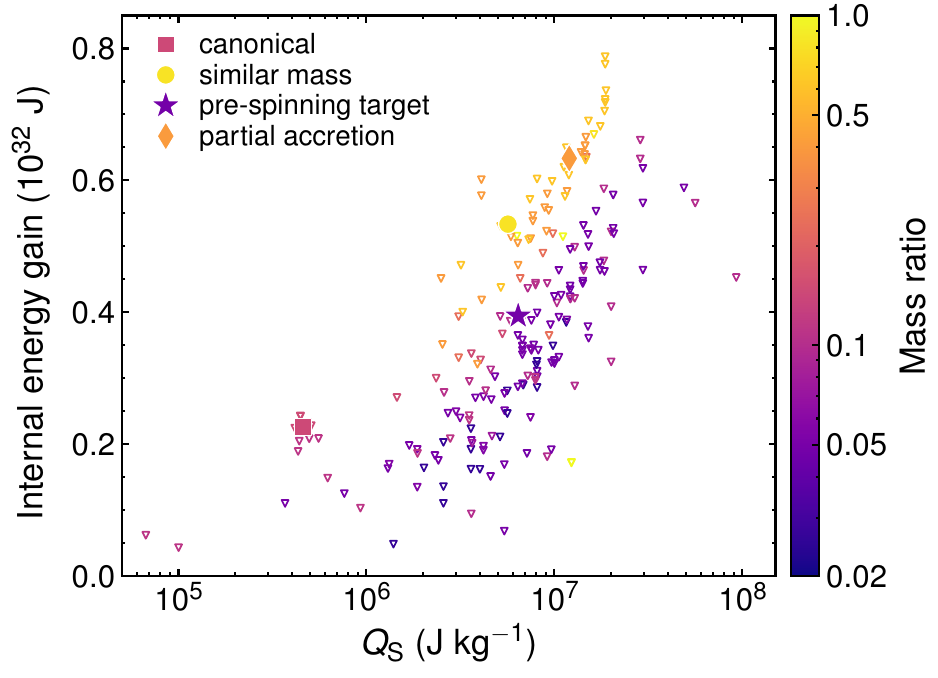}
    \caption{Internal energy gain of the system at 24 hours after the impact as a function of impact energy. The colors of the points indicate the ratio of projectile to target mass. The four example impacts shown throughout this work are indicated by unique symbols.} 
    \label{f:IEgain}
\end{figure}
A common feature across all the giant impacts considered in this work is a significant increase in the total internal energy of the bodies (Figures \ref{f:rawenergy} and \ref{f:ebudgets}). This heating leads to substantial melting and vaporization (see Section \ref{s:heating}). Figure \ref{f:IEgain} shows the gain in internal energy after 24 hours for all simulations in the impact database as a function of modified specific impact energy, $Q_\mathrm{S}$. There is a noticeable general trend of greater internal energy gain with higher impact specific energies and larger mass ratios. {The range of accretional giant impacts that form an Earth-mass planet typically lead to greater gains than for the canonical Moon-forming scenario. The relationship between impact energy and internal energy gain has a complex dependence on the impact geometry as well as the mass ratio; a fit is provided in the supplementary material}.

{For a given specific energy, the largest internal energy gains tend to occur in graze-and-merge impacts. Many of the impacts with the highest energy gains (many of the yellow and orange points in Figure \ref{f:IEgain})} are similar to graze-and-merge impacts in that their cores and mantles `swirl' around each other before merging -- as is the case for the partial accretion example. With larger mass ratios (the {colliding bodies} having more similar masses) the likelihood of graze-and-merge impacts increases \citep{Stewart12}, and we expect that accretional collisions close to the grazing boundary will also share many features with graze-and-merge events.

\subsection{Energy budgets of example giant impacts}

Of the four example impacts, the canonical Moon-forming model has the largest relative contribution to its energy budget from internal energy (Figure \ref{f:ebudgets}a). However, in absolute terms the internal energy gain in the canonical impact is the lowest of the four examples (see Figure \ref{f:rawenergy}). Unsurprisingly, the contribution of kinetic energy is lower in low velocity impacts. With a small impactor (mass ratio $\sim$0.1) the gravitational potential energy change is modest in the canonical model.

The similar mass impactor scenario from \citet{Canup12} has a larger relative contribution from potential energy throughout but still exhibits a significant increase in internal energy (Figure \ref{f:ebudgets}b). Much of the kinetic energy is {exchanged into potential energy after the initial impact, as the bodies separate.} The secondary impact and subsequent swirling of the cores in this graze-and-merge event convert some of this potential {energy} back into kinetic energy, resulting in a fast-spinning post-impact body. This inflated post-impact {body has a less negative potential energy than the minimum during the initial and secondary impacts}. The shock from the second contact generates high velocity ejecta, which holds a significant proportion of the kinetic energy.

The rapidly spinning proto-Earth family of scenarios tend to have a much larger contribution to the energy budget from kinetic energy due to the rotating target and high impact velocity (Figure \ref{f:ebudgets}c). In this example there is no large secondary impact, material raised by the shock that remains bound rapidly falls or compresses back onto the post-impact body and the system approaches equilibrium faster than in the other examples. There is little change in the energy budget beyond 4 hours after the impact. The oscillation between potential and kinetic energy seen between 4 and 9 hours after the impact is due to {spiral arms with locally higher densities} that slowly smooth out. Since the collisions in this scenario result in a fast-spinning body and substantial ejecta, the final energy budget {has} a significant contribution from kinetic energy and from potential energy {gained during} the impact.

The partial accretion impact shares some similarities with each of the previously suggested Moon-forming scenarios. The slower impact leads to a lower relative contribution from kinetic energy, as seen in the canonical model. The relatively massive impactor and substantially inflated body that is left after the impact lead to a large contribution from potential energy, while the lack of a large secondary impact produces the relatively smooth and rapid evolution as seen for the spinning proto-Earth scenario. However, there is much more deformation of the core in this partial accretion scenario as the cores become substantially elongated before they swirl around each other as they merge, similarly to the similar mass impactor scenario. This example accretionary impact shows the largest increase in internal energy, with much of the initial kinetic energy converted into internal and potential energy{. However, the post-impact body has gained significant rotational kinetic energy, leaving it spinning quickly.}


\section{When does potential energy become important?}\label{s:PEcalc}

{In all our giant impact calculations,} a large amount of potential energy is exchanged, and the {post-impact} energy budget {has a significant contribution from} potential energy due to material raised above the original surface of the target, placed into orbit around the remnant, or ejected. {The large contribution of potential energy to giant impacts} is in contrast to the situation for small, cratering impacts in which the potential energy {changes are} considered insignificant, and material ejected from the impact site quickly falls back onto the surface of the body \citep{OKeefe82}. It is important for the study of the consequences of impacts to be able to determine the boundary between these two regimes. {Here we provide an approximation of this transition.}

We can estimate the impactor size at which the potential energy becomes {significant} by considering how the potential energy changes as the impactor penetrates into the target. At the moment of impact the depth of penetration of the impactor into the target, $d$, is zero, and the total potential energy, $E_\mathrm{pot}$, is
\begin{equation}\label{e:PE1}
    E_\mathrm{pot} = - { G M m \over R + r} + E_\mathrm{bind, targ} + E_\mathrm{bind, proj},
\end{equation}
where $G$ is the gravitational constant, $M$ and $m$ are the masses of target and impactor, $R$ and $r$ are the radii of the target and impactor, and $E_\mathrm{bind, targ}$ and $E_\mathrm{bind, proj}$ are the gravitational binding energies of the target and impactor. We note that the binding energies in the combined system are not exactly the same as those of isolated bodies, but we expect the difference to be negligible. If the penetration depth is greater than the impactor radius, the minimum potential energy can be approximated using the expression for potential inside a solid sphere:
\begin{equation}\label{e:PE2}
    E_\mathrm{pot, min} \simeq -  G M m { 3 R^2 - (R + r - d)^2 \over 2 R^3 } + E_\mathrm{bind, targ} + E_\mathrm{bind, proj}.
\end{equation}

From this we obtain the fractional change in potential energy between the start of the impact and the moment of maximum penetration (closest approach of the cores). {This difference in potential energy between the start and the moment of maximum penetration is the basis of the participating potential energy that we defined in section \ref{s:budgets}. The larger this difference in potential energy, the more energy is available to be exchanged during the impact, and affect the impact outcome.} Note that the binding energies are expected to change in the real system, but we do not account for this in our simple estimate.

We can estimate the maximum penetration depth using the impact crater scaling law from \citet{OKeefe93}:
\begin{equation}
    {d \over r } \simeq K { \left( g r \over U^2 \right)^{-{ \mu /( 2 + \mu) }} },
\end{equation}
where $K$ and $\mu$ are empirical constants with values of 1.2 and 0.56 \citep{OKeefe93}, $g$ is the gravitational acceleration, and $U$ is the impact velocity. If we define $U = v \times v_\mathrm{esc}$, where $v_\mathrm{esc}$ is the escape velocity from the target (which is appropriate in the cratering regime, whereas for giant impacts the mutual escape velocity is normally the more appropriate quantity), and substitute for $g$ and $v_\mathrm{esc}$, we obtain,
\begin{equation}\label{e:depth}
    {d \over r } \simeq K v^{ 2\mu /( 2 + \mu) } { \left( r \over 2 R \right)^{-{ \mu /( 2 + \mu) }} }.
\end{equation}
\begin{figure}
    \centering
    \includegraphics[width=\columnwidth]{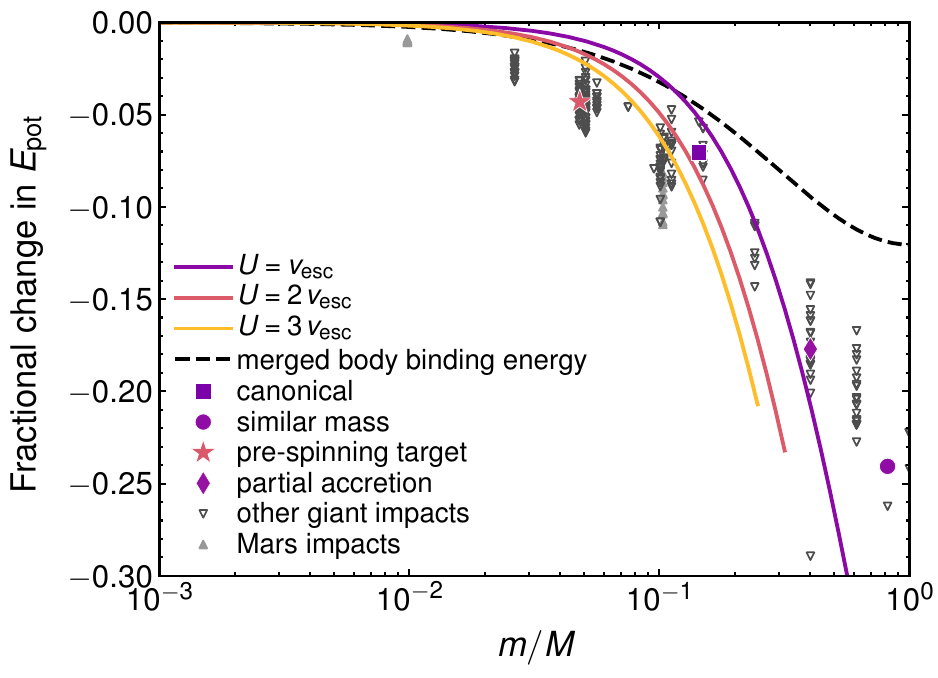}
    \caption{Estimated fractional change in gravitational potential energy between the start of the impact and the point of potential minimum due to penetration of the impactor into the target as a function of impactor to target mass ratio. The solid lines indicate the release of potential energy according to the cratering estimate, with colors showing different impact velocities. The dashed line indicates the fractional difference between initial potential energy and the binding energy of a single body produced by perfectly merging the {colliding bodies} while maintaining a constant density {and zero angular momentum}. The fractional change in potential energy between the start and potential minimum for the impact simulations are also shown (dark grey triangles), with the four example impacts colored according to scaled impact velocity. {Impacts onto Mars-mass bodies taken from \citet{Carter18} are also shown (filled light grey triangles); note that these generally have impact velocities that are larger multiples of the escape velocity than for the Earth-mass giant impacts.} The model lines terminate where the crater depth equals the radius of the target.}
    \label{f:PEvsa}
\end{figure}
We can thus relate the fractional change in potential energy due to penetration of the impactor into the target to the size of the impactor, and the impact velocity, using equations \ref{e:PE1}, \ref{e:PE2} and \ref{e:depth}. The result is shown in Figure \ref{f:PEvsa}, along with the changes in potential energy between the start and the potential minimum from our impact simulations {and the Mars-mass impact simulations from \citet{Carter18}}. Faster impacts lead to greater penetration and, thus, greater release of potential energy for a given impactor size. As the penetration depth calculated using equation \ref{e:depth}, which is based on impacts into a half-space, approaches unity or the size of the impactor approaches the size of the target, this simple model loses validity. Figure \ref{f:PEvsa} also shows the predicted change in potential energy caused by the merging of two uniform density spherical {bodies} into a single spherical body of the same density (black dashed line). Note that the change in potential energy during impacts exceeds this estimate in all cases.

Given the simplicity of this crater depth calculation and the variation in {the impact angle and} the oblateness and densities of the {colliding bodies} in the simulations, the match {with the simulation data} is {fairly} good. {The penetration depth scaling captures much of the increase in potential energy change with increasing mass ratio, although it fails as the mass ratio approaches unity.} From this simple estimate we see that the potential energy released during the impact becomes significant at impactor sizes above approximately 1\% of the target mass, with only a minor dependence on impact velocity. There is no strict definition for a giant impact; one way to differentiate a giant impact from a cratering event is when potential energy has a substantial role in the energy budget of the event. For an Earth-sized body, the above estimate implies potential energy could be a {key} component of the energy budget for impactors with radii larger than $\sim1500\,$km.


\section{Heating and thermal state in giant impacts}\label{s:heating}

\begin{figure*}
    \centering
    \includegraphics[width=1.\textwidth]{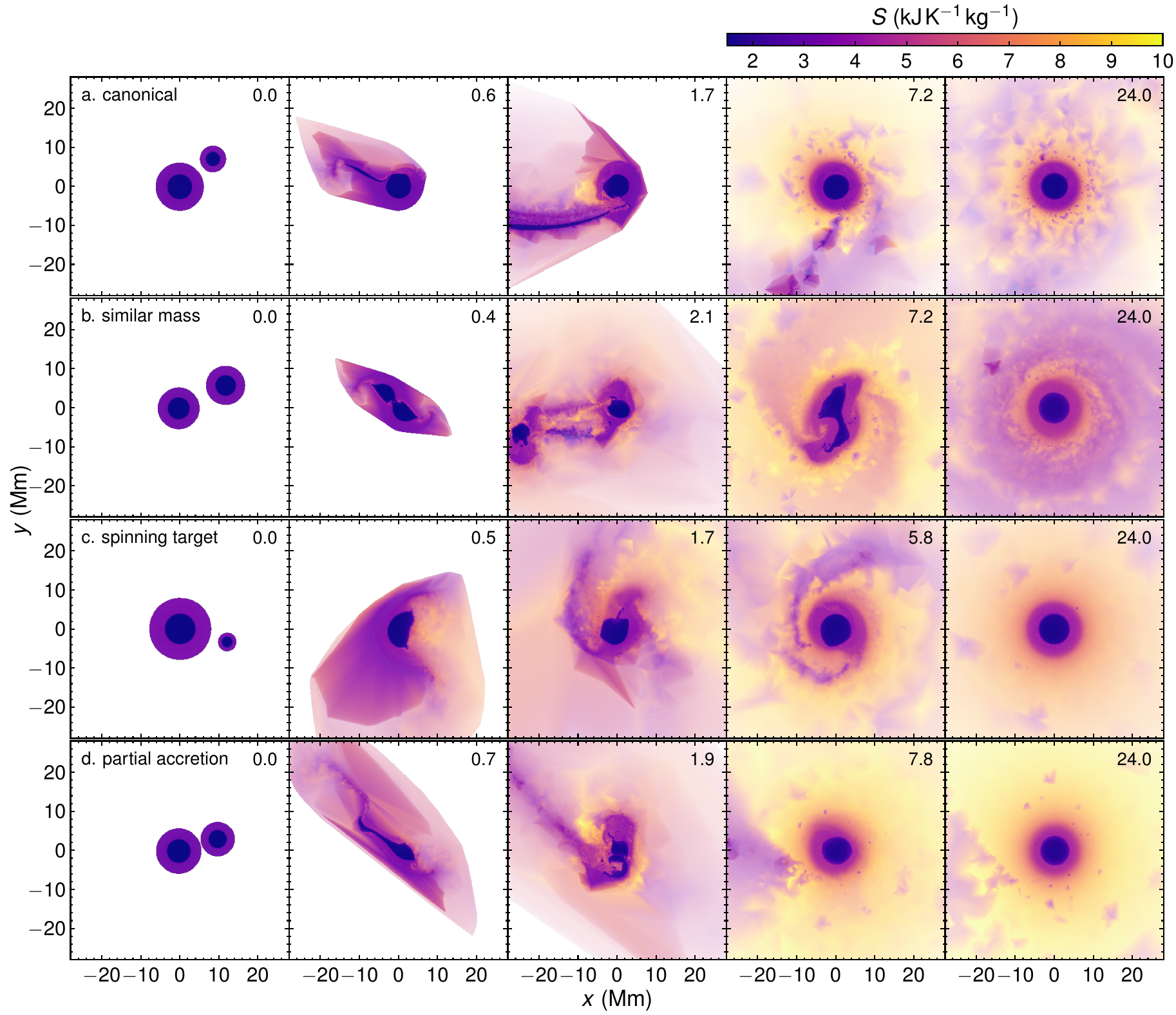}
    \caption{Time sequences of impacts showing entropy in the equatorial plane in SPH simulations (as for the density profiles shown in Figure \ref{f:impactseq}). The color scale indicates the entropy of the material, and the density is indicated by the transparency, where the lowest density material is almost entirely transparent. (a) A canonical Moon-forming impact; (b) a similar mass impactors Moon-forming scenario; (c) a spinning proto-Earth Moon-forming impact; and (d) a partial accretion impact. The time in hours is shown in the top right corner of each panel. Images are recentered on the gravitational potential minimum of the system in each panel. Accompanying animations are supplied in the supplementary materials.}
    \label{f:entropy}
\end{figure*}
Giant impacts cause a significant increase in the specific entropy of the mantle. We examine entropy because it is not affected by adiabatic decompression and as such indicates the thermal state independent of the pressure the material is under. This increase in entropy of the bodies tracks the increases in internal energy. There is some entropy gain during the initial impact, but a larger increase occurs after maximum decompression as displaced material falls or compresses onto the remnant body. The secondary impacts seen in Figure \ref{f:ebudgets} are not associated with a significant increase in entropy, it is the re-impacting `streams' of material (see Figure \ref{f:entropy}) that are responsible for the gradual entropy gain.

\begin{figure}
    \centering
    \includegraphics[width=\columnwidth]{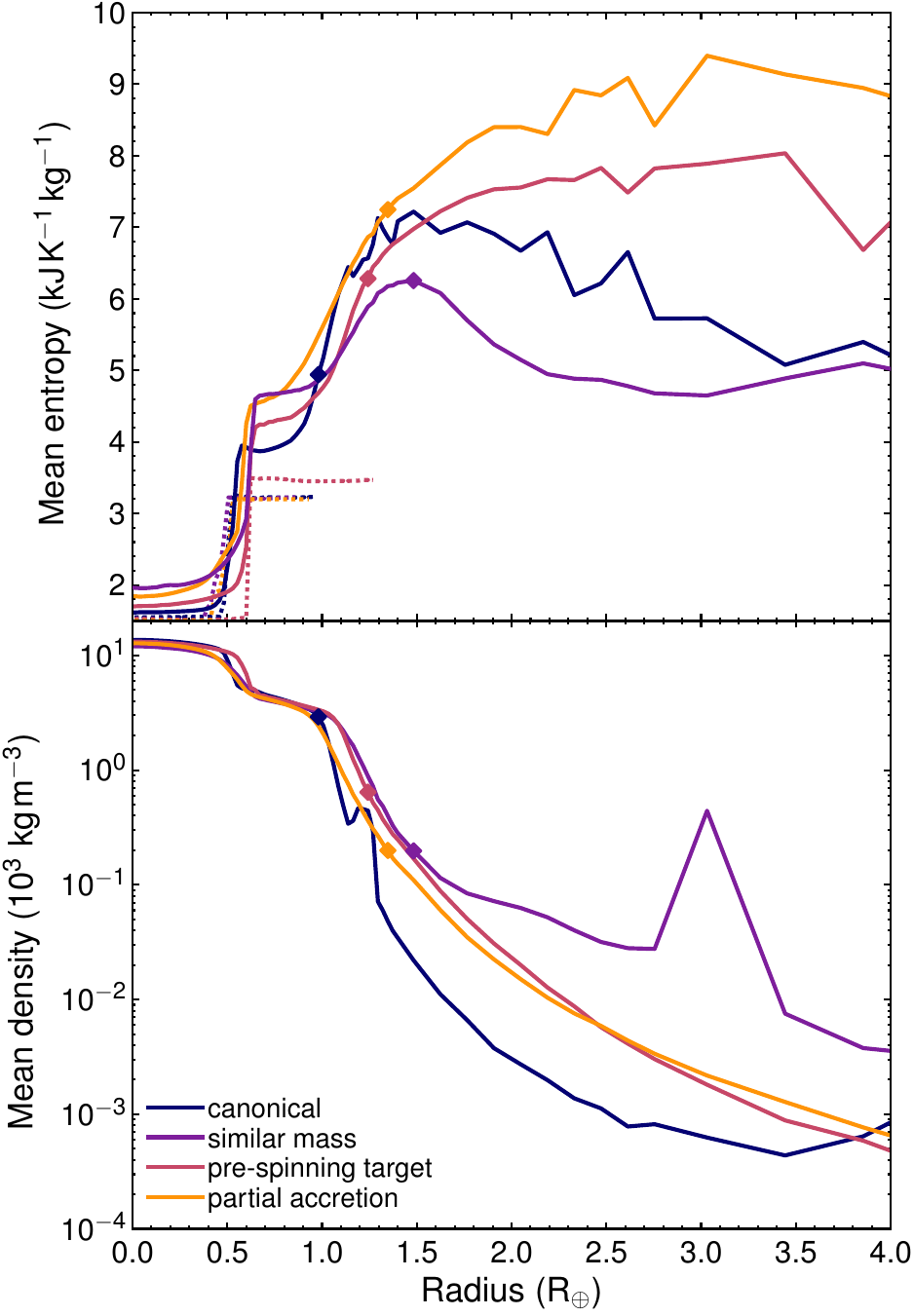}
    \caption{Midplane specific entropy and density profiles of the post-impact bodies at 24 hours. The dotted lines indicate the initial entropy profiles of the targets. Diamonds indicate the equatorial radius within which 95\% of the mass is enclosed. The density peaks at $\sim$1.2\,R$_\oplus$ for the canonical example and at $\sim$3.1\,R$_\oplus$ for the similar mass impactors example are due to reaccreting and orbiting clumps (visible in the upper left quadrant of the final panels in Figure \ref{f:impactseq}).}
    \label{f:entropyprofile}
\end{figure}

The conversion of kinetic and potential energies to internal energy and associated increases in specific entropy are non-uniform (see Figures \ref{f:entropyprofile} and \ref{f:entropyevolution}). The iron core exhibits a smaller increase in entropy than the mantle, {but within both core and mantle the entropy gain is heterogeneous.} In the remainder of this section we will examine the specific entropy of the mantle and core separately and discuss the implications of their thermal histories.

\subsection{Mantle heating}

\begin{figure*}
    \centering
    \includegraphics[width=1.\textwidth]{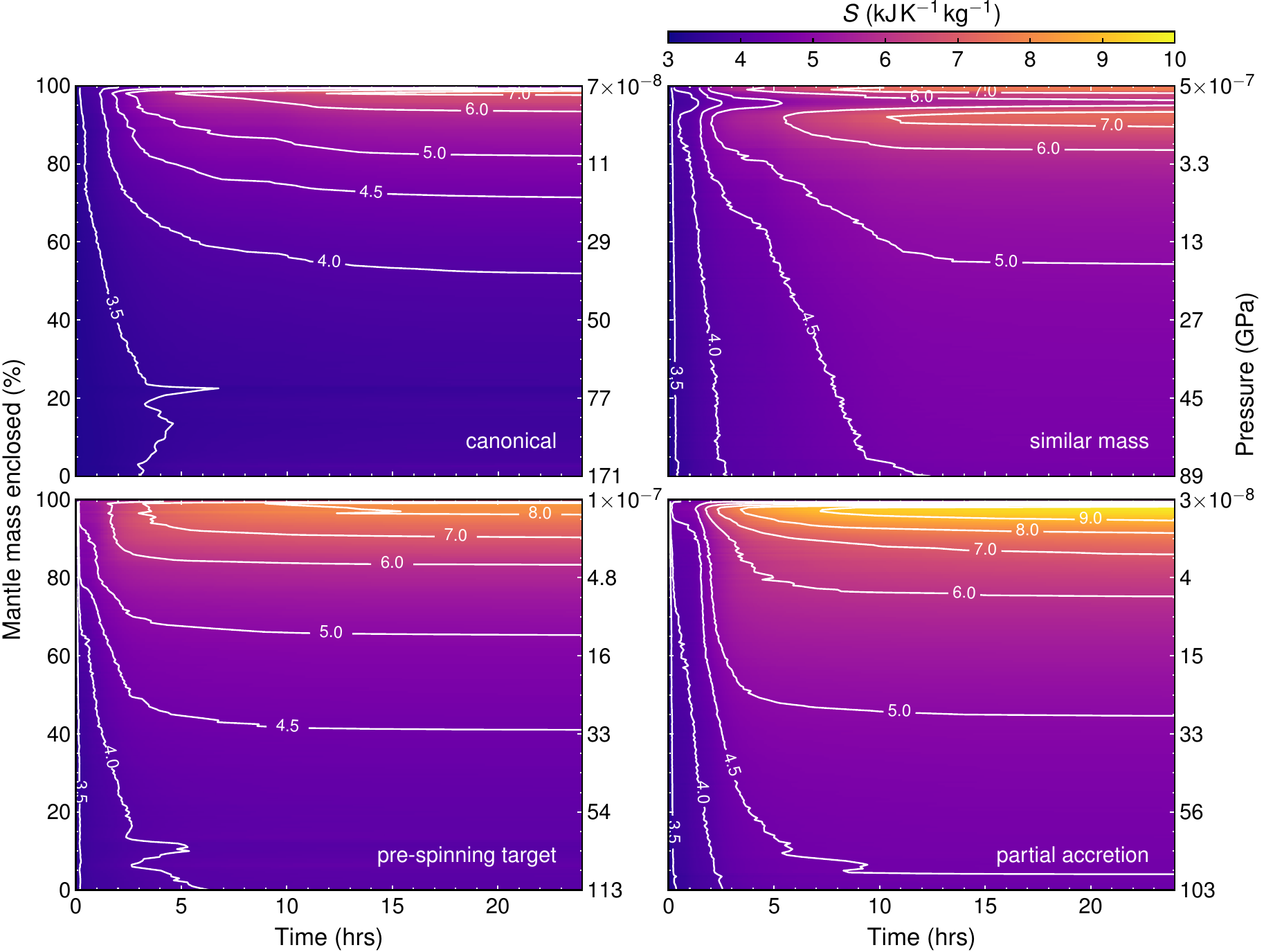}
    \caption{{Mean} specific entropy profiles of the mantle as a function of pressure within the post-impact body, plotted as mass fraction of bound mantle, as the collision proceeds and the post-impact body equilibrates. Entropies are averaged for all parcels of mass at a given pressure. {Pressures are those within the post-impact body at 24 hours, the low pressures at the core-mantle boundary compared to present-day Earth are a real feature of the post-impact bodies, as discussed in \citet{Lock19a}.} The entropy of material is shown according to it's location within the post-impact body at 24 hours. {Contours are shown at specific entropies of 3.5, 4, 4.5, 5, 6, 7, 8 and 9\,kJ\,K$^{-1}$\,kg$^{-1}$.} Post-impact the mantle is stratified in all the examples, with the hottest material in the upper layers of the body or the surrounding disk-like region. Some of the impacts show a lower entropy band between 95 and 100\% 'enclosed' mass, this is due to low entropy material that is still (re-)accreting, or orbiting within the disk-like regions. As for internal energy, the bulk of the entropy increase does not coincide with the initial impact, but occurs over a longer timescale during gravitational reequilibration and accretion of debris.}
    \label{f:entropyevolution}
\end{figure*}

The lower mantle experiences a moderate increase in specific entropy ($\Delta S \lesssim 1.5$\,kJ\,K$^{-1}$\,kg$^{-1}$), which varies considerably between the different impact scenarios {(see Figure \ref{f:entropyprofile})}, whilst the outer layers and ejecta are substantially heated by the re-impacting and re-equilibrating material. The release of potential energy during the {infall and reequilibration} of the post-impact body generates secondary shocks which cause the outer silicate material to vaporize, as indicated by the orange and yellow material in Figure \ref{f:entropy} ($S \gtrsim 7$\,kJ\,K$^{-1}$\,kg$^{-1}$).

{After a giant impact, the post-impact body has a highly stratified mantle}. Figure \ref{f:entropyevolution} shows the entropy structure of bound mantle mass as a function of pressure within the post-impact body. The lower mantle experiences relatively little heating while the upper layers reach high entropy (7--10\,kJ\,K$^{-1}$\,kg$^{-1}$), largely due to heating that occurs after the initial impact (after the first hour). Note that some colder material can be rotationally supported within the orbiting disk-like regions, which is partially responsible for the bands of lower entropy (purple, $<$6\,kJ\,K$^{-1}$\,kg$^{-1}$) material at low pressures in Figure \ref{f:entropyevolution}. Moonlets orbiting in the disk-like regions are also likely to be lower entropy than much of the disk, but their higher internal pressures cause them to be indistinguishable in Figure \ref{f:entropyevolution} from slightly deeper layers that are dominated by higher entropy material (orange, 7--9\,kJ\,K$^{-1}$\,kg$^{-1}$).

We see very similar specific entropy gains in the mantle to that shown by \citet{Nakajima15}. The entropy increase through most of the mantle is larger for higher energy, higher angular momentum impacts compared to the canonical scenario. {Using the same entropy change melting criterion used by \citet{Nakajima15} and their assumption of starting at the solidus, $\Delta S>0.623$\,kJ\,K$^{-1}$\,kg$^{-1}$, we find similar mass fractions of the mantle melt as a result of the impact (62\% for the canonical example, and 100\% for the other three). However, it is important to recall that these simulations do not take into account material strength, which can significantly alter the temperature changes and fraction of material that melts during impacts \citep[see][]{Emsenhuber18}. The effect of material strength on melting is particularly important at later times, when the majority of the heating occurs, as the fluid assumption may become less valid. Future work should address the final melt fraction using an equation of state that incorporates the melt curve, considerations for the range of possible initial temperatures, and a code that includes material strength.}

The inner region of the canonical Moon-forming disk has a high entropy ($S \sim 7$\,kJ\,K$^{-1}$\,kg$^{-1}$), so much of the mass is in the vapor phase, and there is a clear distinction between the planet and the disk. Higher energy, higher angular momentum impact scenarios lead to hotter mantles, and continuous, extended structures known as synestias \citep{Lock17}. These vapor-dominated structures have shallower entropy and (surface) density gradients than seen in the post-impact body for the canonical scenario (see Figure \ref{f:entropyprofile}). The disk-like regions of synestias are vapor-dominated to much greater distances from the center of the post-impact body than the canonical disk.

The outer layers of the mantle experience significant heating via secondary shocks when this material is at lower densities and thus more compressible. The lower mantle is most sensitive to the initial impact shock and can experience relatively little disturbance and heating in a wide range of impact scenarios (see Figure \ref{f:entropyevolution}). Weak heating of the lower mantle due to the shock could leave parts of the lower mantle partially solid. SPH methods are not expected to perfectly capture the mixing and redistribution of mantles due to the hydrodynamic rheology. However, in general, the heating of the outer mantle occurs sufficiently late that we would not expect substantial mixing between these heated outer layers and the cooler lower mantle.

{\citet{Nakajima15} pointed out that shearing flows within the stratified mantle may become unstable and mix. They considered a simple energy balance of the bulk kinetic energy difference and the bulk potential energy difference between the upper and lower mantle between the start and end of the impact. Our work shows that these terms vary substantially in time and are heterogeneous in space. The strongest shear involving the lower mantle is near the beginning of the event (Figure \ref{f:ebudgets}) and the strong stratification between the lower and upper mantle develops later in time (Figure \ref{f:entropyevolution}). In addition, giant impacts commonly produce synestias. A synestia has a cororating inner region and differentially rotating disk region. The late-time differential rotation occurs in the disk region of the synestia (which can hold a substantial amount of the mass; up to 15\% in our simulations). The disk may mix via instabilities driven by the shearing flow, but the low-entropy lower mantle lies below a higher-entropy upper mantle within the corotating region. It should also be noted that there is also a strong angular momentum barrier to mixing perpendicularly to the rotation axis. Thus, the processes that heat the outer regions of the synestia (upper mantle and disk) late in the impact (e.g.\ assimilation of infalling debris) are not expected to lead to mixing of the entire lower mantle.}

{Several studies \citep[e.g.][]{Mukhopadhyay12,Graham16,Rizo16,Mundl17,Williams19} have shown evidence that some regions of the Earth's mantle (likely deep in the mantle) exhibit primordial isotopic signatures, suggesting that they have been preserved since accretion, and likely survived the Moon-forming impact. The entropy changes discussed above suggest that the entire mantle melts. However, melting alone is not a sufficient condition for destruction of primitive mantle reservoirs because mixing is inhibited by the entropy stratification and angular momentum exchange. The entropy gain in a strengthless impact simulation is not sufficient to rule out high energy, high angular momentum impacts for the origin of the Moon.}

\subsection{Core heating}

\begin{figure*}
    \centering
    \includegraphics[width=1.\textwidth]{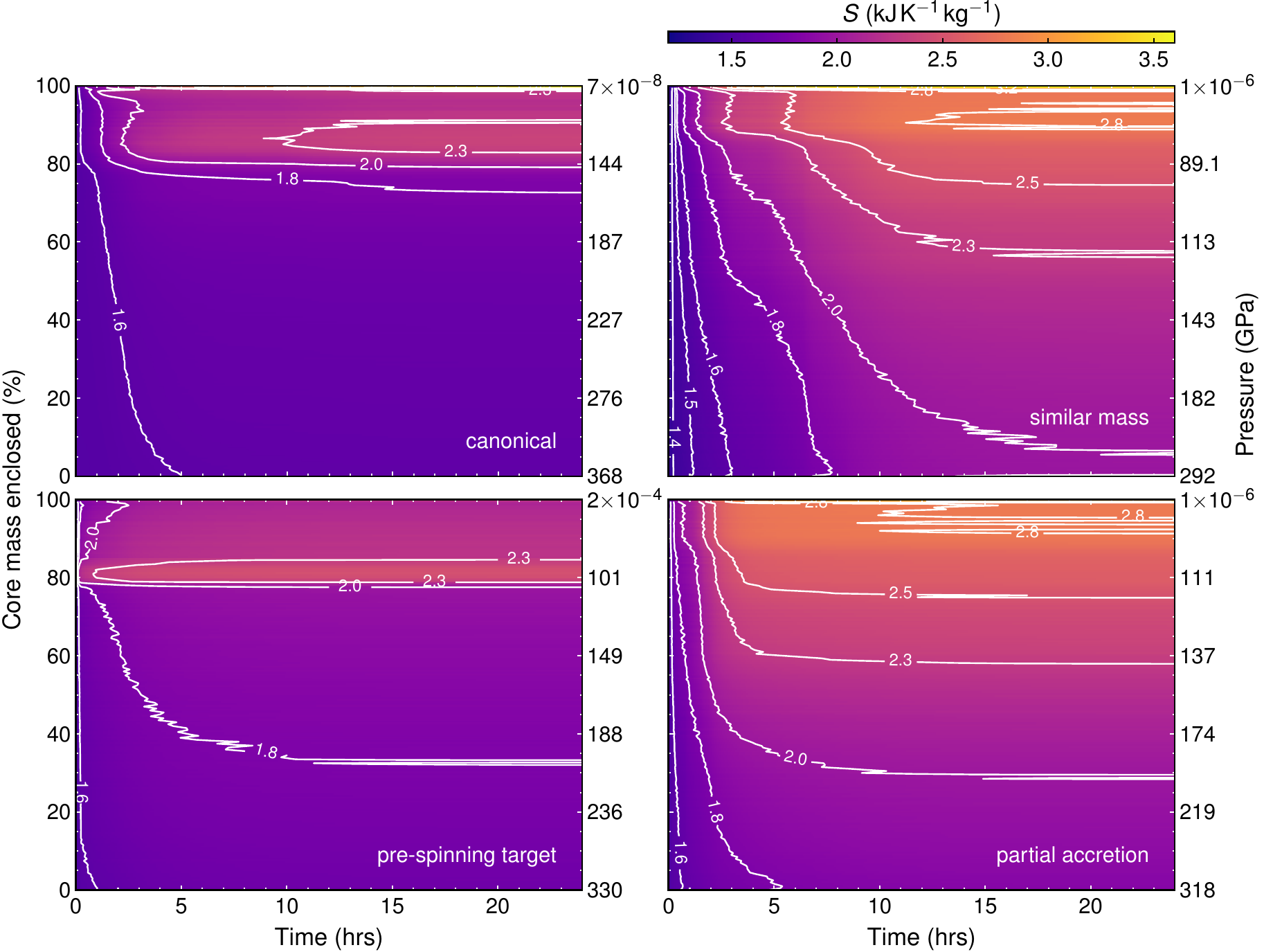}
    \caption{{Mean specific entropy profiles of the core as a function of pressure within the post-impact body, plotted as mass fraction of bound iron, as the collision proceeds and the post-impact body equilibrates. Entropies are averaged for all parcels of mass at a given pressure. Note that some iron particles are located within the mantle and extended atmosphere at 24 hours, giving rise to the low pressures for 100\% 'enclosed' mass. The entropy of material is shown according to it's location within the post-impact body at 24 hours. Contours are shown at specific entropies of 1.4, 1.5, 1.6, 1.8, 2, 2.3, 2.5, 2.8 and 3.2\,kJ\,K$^{-1}$\,kg$^{-1}$. Post-impact the core is stratified in all the examples.}}
    \label{f:coreentropy}
\end{figure*}
Giant impacts also heat the cores of the {colliding bodies} (see Figures \ref{f:entropyprofile} {and \ref{f:coreentropy}}). The core experiences greater heating in the similar mass impactor scenario and the partial accretion example due to the substantial deformation of the cores of both {bodies} during the collision (see  Figures \ref{f:impactseq} and \ref{f:entropy}). {The core experiences greater deformation in events with more similarly sized bodies or smaller impact parameters.} The merged cores {in the final} post-impact bodies can also be thermally stratified. The higher entropy (2.5--3\,kJ\,K$^{-1}$\,kg$^{-1}$) of the outer region of the core of the post-impact body is particularly noticeable for the partial accretion example (the mean entropy exhibits a substantial increase between 0.4 and 0.55 R$_\oplus$, Figures \ref{f:entropyprofile} {and \ref{f:coreentropy}}).

In the example impacts shown here, the impactor and target cores exhibit a diverse range of phenomena. In the canonical impact the core of the target experiences very little deformation, but the impactor's core is substantially disrupted, with many fragments of iron falling through the post-impact body's mantle over the entire midplane circumference. The canonical impact results in little heating of the {target's} core. In the similar-mass impactor scenario the cores are highly deformed, leading to a greater increase in entropy, but there is less opportunity for equilibration between the core and the target mantle as much of the core remains as a continuous mass. The spinning proto-Earth scenario shows the least core deformation of all the examples, with the core of the small impactor falling straight through the target's mantle, and only slight deformation as the post-impact body forms and relaxes. The example partial accretion impact shows substantial deformation and heating of the cores of target and impactor, with large {fragments} being `flung' out into the flowing mantle, though this phase is short-lived.

It is important to note, however, that these SPH simulations may not accurately capture the dynamics of the core if it undergoes significant deformation or disruption. In cases where the cores are disrupted, SPH methods do not treat the breakup or multiphase flow accurately, and the formation of fragments may be substantially different in reality. This is due to the Lagrangian nature of SPH, in which mass cannot be exchanged between particles; the lack of sufficient resolution to resolve all instabilities; and the inhibition of mixing and shear by the smoothing kernel.

{Without substantial deformation or heating of the target core, mixing may be inhibited, preserving any existing density structure within the core. This may inhibit convection and the generation of a dynamo. Strong deformation, on the other hand, may lead to mixing between the target and impactor cores that can erase density structures in the core and lead to global core convection and a strong dynamo as expected for the Earth \citep{Jacobson17}}.

These example impacts show that giant impacts may have very different degrees of equilibration between core and mantle reservoirs, and that this equilibration could be very sensitive to the parameters of each impact, not just the impact energy. {Metal equilibration could range from negligible to near-total in different impacts, but may also be affected by material strength \citep{Emsenhuber18}.} One should not use a single core mantle equilibration factor for all giant impacts as often assumed when modeling metal silicate equilibration \citep[e.g.][]{Nimmo10,Rudge10}. Our results also suggest large differences in the degree of heating of terrestrial planet cores in giant impacts. We also note that different parts of a core will likely equilibrate with the mantle at different temperatures and pressures depending on the core's thermal history during the impact. {The degree and conditions of core equilibration are crucial to our understanding of accretion, and more detailed work is required to track the variable conditions of metal-silicate equilibration during giant impacts to inform geochemical models.}


\section{Conclusions}

The energies involved in giant impacts are extremely large, and every impact is different. There are substantial differences in the evolution of the energy budgets between different giant impacts, and significant differences in the heating of the post-impact body. The wide variation in energy evolution between different impacts means that one should not take a single impact (e.g.\ the canonical Moon-forming impact) as being characteristic of all giant impacts.

Giant impacts lead to substantial increases in the internal energies of bodies, due to the release of kinetic and potential energy. There is a general trend of greater internal energy gain with higher impact specific energies and larger mass ratios, and it is clear that the impacts in our database typically lead to greater gains than for the canonical Moon-forming scenario (see Figure \ref{f:IEgain}). 

The distribution of energy after a giant impact is crucial for understanding the amount of melting and vaporization, and the thermal state of the final planet. The variation in the degree of thermal inflation and spin rate induced by an impact will cause differences in the later evolution of the planet {\citep{Chau18,Lock19}}. {The variation in energy budgets between giant impacts also has important implications for metal-silicate equilibration and core evolution.}

\bibliographystyle{aasjournal}
\bibliography{references}

\begin{thebibliography}{}
\expandafter\ifx\csname natexlab\endcsname\relax\def\natexlab#1{#1}\fi
\providecommand{\url}[1]{\href{#1}{#1}}
\providecommand{\dodoi}[1]{doi:~\href{http://doi.org/#1}{\nolinkurl{#1}}}
\providecommand{\doeprint}[1]{\href{http://ascl.net/#1}{\nolinkurl{http://ascl.net/#1}}}
\providecommand{\doarXiv}[1]{\href{https://arxiv.org/abs/#1}{\nolinkurl{https://arxiv.org/abs/#1}}}

\bibitem[{{Asphaug}(2014)}]{Asphaug14}
{Asphaug}, E. 2014, Annual Review of Earth and Planetary Sciences, 42, 551,
  \dodoi{10.1146/annurev-earth-050212-124057}

\bibitem[{{Asphaug} {et~al.}(2006){Asphaug}, {Agnor}, \&
  {Williams}}]{Asphaug06}
{Asphaug}, E., {Agnor}, C.~B., \& {Williams}, Q. 2006, \nat, 439, 155,
  \dodoi{10.1038/nature04311}

\bibitem[{{Barr}(2016)}]{Barr16}
{Barr}, A.~C. 2016, Journal of Geophysical Research (Planets), 121, 1573,
  \dodoi{10.1002/2016JE005098}

\bibitem[{{Benz} {et~al.}(2007){Benz}, {Anic}, {Horner}, \& {Whitby}}]{Benz07}
{Benz}, W., {Anic}, A., {Horner}, J., \& {Whitby}, J.~A. 2007, \ssr, 132, 189,
  \dodoi{10.1007/s11214-007-9284-1}

\bibitem[{{Cameron} \& {Ward}(1976)}]{Cameron+Ward}
{Cameron}, A.~G.~W., \& {Ward}, W.~R. 1976, in Lunar and Planetary Science
  Conference, Vol.~7, 120

\bibitem[{{Canup}(2004)}]{Canup04}
{Canup}, R.~M. 2004, \icarus, 168, 433, \dodoi{10.1016/j.icarus.2003.09.028}

\bibitem[{{Canup}(2005)}]{Canup05}
---. 2005, Science, 307, 546, \dodoi{10.1126/science.1106818}

\bibitem[{{Canup}(2008)}]{Canup08}
---. 2008, \icarus, 196, 518, \dodoi{10.1016/j.icarus.2008.03.011}

\bibitem[{{Canup}(2012)}]{Canup12}
---. 2012, Science, 338, 1052, \dodoi{10.1126/science.1226073}

\bibitem[{{Canup} \& {Asphaug}(2001)}]{Canup01}
{Canup}, R.~M., \& {Asphaug}, E. 2001, \nat, 412, 708

\bibitem[{{Canup} \& {Righter}(2000)}]{Canup00}
{Canup}, R.~M., \& {Righter}, K. 2000, {Origin of the earth and moon}, ed.
  R.~M. {Canup} \& K.~{Righter} (Tucson: University of Arizona Press)

\bibitem[{{Carter} {et~al.}(2018){Carter}, {Leinhardt}, {Elliott}, {Stewart},
  \& {Walter}}]{Carter18}
{Carter}, P.~J., {Leinhardt}, Z.~M., {Elliott}, T., {Stewart}, S.~T., \&
  {Walter}, M.~J. 2018, Earth and Planetary Science Letters, 484, 276,
  \dodoi{10.1016/j.epsl.2017.12.012}

\bibitem[{Carter {et~al.}(2019)Carter, Lock, \& Stewart}]{Carter19GIDATA}
Carter, P.~J., Lock, S.~J., \& Stewart, S.~T. 2019, {Replication Data for:
  ``The energy budgets of giant impacts''}, V1,  Harvard Dataverse,
  \dodoi{10.7910/DVN/YYNJSX}

\bibitem[{{Chambers} \& {Wetherill}(1998)}]{Chambers98}
{Chambers}, J.~E., \& {Wetherill}, G.~W. 1998, \icarus, 136, 304,
  \dodoi{10.1006/icar.1998.6007}

\bibitem[{{Chau} {et~al.}(2018){Chau}, {Reinhardt}, {Helled}, \&
  {Stadel}}]{Chau18}
{Chau}, A., {Reinhardt}, C., {Helled}, R., \& {Stadel}, J. 2018, \apj, 865, 35,
  \dodoi{10.3847/1538-4357/aad8b0}

\bibitem[{{{\'C}uk} \& {Stewart}(2012)}]{Cuk12}
{{\'C}uk}, M., \& {Stewart}, S.~T. 2012, Science, 338, 1047,
  \dodoi{10.1126/science.1225542}

\bibitem[{{Dauphas}(2017)}]{Dauphas17}
{Dauphas}, N. 2017, \nat, 541, 521, \dodoi{10.1038/nature20830}

\bibitem[{{Dauphas} {et~al.}(2014){Dauphas}, {Burkhardt}, {Warren}, \&
  {Teng}}]{Dauphas14}
{Dauphas}, N., {Burkhardt}, C., {Warren}, P., \& {Teng}, F.-Z. 2014,
  Philosophical Transactions of the Royal Society of London Series A, 372,
  2013.0244, \dodoi{10.1098/rsta.2013.0244}

\bibitem[{{Emsenhuber} \& {Asphaug}(2019)}]{Emsenhuber19}
{Emsenhuber}, A., \& {Asphaug}, E. 2019, \apj, 875, 95,
  \dodoi{10.3847/1538-4357/ab0c1d}

\bibitem[{{Emsenhuber} {et~al.}(2018){Emsenhuber}, {Jutzi}, \&
  {Benz}}]{Emsenhuber18}
{Emsenhuber}, A., {Jutzi}, M., \& {Benz}, W. 2018, \icarus, 301, 247,
  \dodoi{10.1016/j.icarus.2017.09.017}

\bibitem[{{Genda} {et~al.}(2012){Genda}, {Kokubo}, \& {Ida}}]{Genda12}
{Genda}, H., {Kokubo}, E., \& {Ida}, S. 2012, \apj, 744, 137,
  \dodoi{10.1088/0004-637X/744/2/137}

\bibitem[{Graham {et~al.}(2016)Graham, Michael, \& Shea}]{Graham16}
Graham, D.~W., Michael, P.~J., \& Shea, T. 2016, Earth and Planetary Science
  Letters, 454, 192 , \dodoi{https://doi.org/10.1016/j.epsl.2016.09.016}

\bibitem[{{Hartmann} \& {Davis}(1975)}]{Hartmann+Davis}
{Hartmann}, W.~K., \& {Davis}, D.~R. 1975, \icarus, 24, 504,
  \dodoi{10.1016/0019-1035(75)90070-6}

\bibitem[{{Hollyday} {et~al.}(2017){Hollyday}, {Stewart}, {Leinhardt},
  {Carter}, \& {Lock}}]{Hollyday17}
{Hollyday}, G.~O., {Stewart}, S.~T., {Leinhardt}, Z.~M., {Carter}, P.~J., \&
  {Lock}, S.~J. 2017, in Lunar and Planetary Science Conference, Vol.~48, 48th
  Lunar and Planetary Science Conference, 2606

\bibitem[{{Hosono} {et~al.}(2019){Hosono}, {Karato}, {Makino}, \&
  {Saitoh}}]{Hosono19}
{Hosono}, N., {Karato}, S.-i., {Makino}, J., \& {Saitoh}, T.~R. 2019, Nature
  Geoscience, 12, 418, \dodoi{10.1038/s41561-019-0354-2}

\bibitem[{{Jackson} \& {Wyatt}(2012)}]{Jackson12}
{Jackson}, A.~P., \& {Wyatt}, M.~C. 2012, \mnras, 425, 657,
  \dodoi{10.1111/j.1365-2966.2012.21546.x}

\bibitem[{{Jacobson} {et~al.}(2017){Jacobson}, {Rubie}, {Hernlund},
  {Morbidelli}, \& {Nakajima}}]{Jacobson17}
{Jacobson}, S.~A., {Rubie}, D.~C., {Hernlund}, J., {Morbidelli}, A., \&
  {Nakajima}, M. 2017, Earth and Planetary Science Letters, 474, 375,
  \dodoi{10.1016/j.epsl.2017.06.023}

\bibitem[{{Korycansky} {et~al.}(1990){Korycansky}, {Bodenheimer}, {Cassen}, \&
  {Pollack}}]{Korycansky90}
{Korycansky}, D.~G., {Bodenheimer}, P., {Cassen}, P., \& {Pollack}, J.~B. 1990,
  \icarus, 84, 528, \dodoi{10.1016/0019-1035(90)90051-A}

\bibitem[{{Leinhardt} \& {Stewart}(2012)}]{Leinhardt12}
{Leinhardt}, Z.~M., \& {Stewart}, S.~T. 2012, \apj, 745, 79,
  \dodoi{10.1088/0004-637X/745/1/79}

\bibitem[{{Lock}(2019)}]{HERCULES}
{Lock}, S.~J. 2019, {sjl499/HERCULESv1\_user: HERCULES planetary structure
  code}, v1.0.0,  Zenodo, \dodoi{10.5281/zenodo.3509365}

\bibitem[{{Lock} \& {Stewart}(2017)}]{Lock17}
{Lock}, S.~J., \& {Stewart}, S.~T. 2017, Journal of Geophysical Research
  (Planets), 122, 950, \dodoi{10.1002/2016JE005239}

\bibitem[{Lock \& Stewart(2019)}]{Lock19a}
Lock, S.~J., \& Stewart, S.~T. 2019, Science Advances, 5,
  \dodoi{10.1126/sciadv.aav3746}

\bibitem[{{Lock} {et~al.}(2019){Lock}, {Stewart}, \& {{\'C}uk}}]{Lock19}
{Lock}, S.~J., {Stewart}, S.~T., \& {{\'C}uk}, M. 2019, Earth and Planetary
  Science Letters, arXiv:1910.00619.
\newblock \doarXiv{1910.00619}

\bibitem[{{Lock} {et~al.}(2018){Lock}, {Stewart}, {Petaev}, {Leinhardt},
  {Mace}, {Jacobsen}, \& {Cuk}}]{Lock18}
{Lock}, S.~J., {Stewart}, S.~T., {Petaev}, M.~I., {et~al.} 2018, Journal of
  Geophysical Research (Planets), 123, 910, \dodoi{10.1002/2017JE005333}

\bibitem[{{Marcus} {et~al.}(2009){Marcus}, {Stewart}, {Sasselov}, \&
  {Hernquist}}]{Marcus09}
{Marcus}, R.~A., {Stewart}, S.~T., {Sasselov}, D., \& {Hernquist}, L. 2009,
  \apjl, 700, L118, \dodoi{10.1088/0004-637X/700/2/L118}

\bibitem[{{Marinova} {et~al.}(2008){Marinova}, {Aharonson}, \&
  {Asphaug}}]{Marinova08}
{Marinova}, M.~M., {Aharonson}, O., \& {Asphaug}, E. 2008, \nat, 453, 1216,
  \dodoi{10.1038/nature07070}

\bibitem[{{Melosh}(2007)}]{Melosh07}
{Melosh}, H.~J. 2007, Meteoritics and Planetary Science, 42, 2079,
  \dodoi{10.1111/j.1945-5100.2007.tb01009.x}

\bibitem[{Mukhopadhyay(2012)}]{Mukhopadhyay12}
Mukhopadhyay, S. 2012, Nature, 486, 101 EP .
\newblock \url{https://doi.org/10.1038/nature11141}

\bibitem[{{Mundl} {et~al.}(2017){Mundl}, {Touboul}, {Jackson}, {Day}, {Kurz},
  {Lekic}, {Helz}, \& {Walker}}]{Mundl17}
{Mundl}, A., {Touboul}, M., {Jackson}, M.~G., {et~al.} 2017, Science, 356, 66,
  \dodoi{10.1126/science.aal4179}

\bibitem[{{Nakajima} \& {Stevenson}(2015)}]{Nakajima15}
{Nakajima}, M., \& {Stevenson}, D.~J. 2015, Earth and Planetary Science
  Letters, 427, 286, \dodoi{10.1016/j.epsl.2015.06.023}

\bibitem[{Nimmo {et~al.}(2010)Nimmo, O'Brien, \& Kleine}]{Nimmo10}
Nimmo, F., O'Brien, D., \& Kleine, T. 2010, Earth and Planetary Science
  Letters, 292, 363 , \dodoi{https://doi.org/10.1016/j.epsl.2010.02.003}

\bibitem[{{O'Keefe} \& {Ahrens}(1982)}]{OKeefe82}
{O'Keefe}, J.~D., \& {Ahrens}, T.~J. 1982, \jgr, 87, 6668,
  \dodoi{10.1029/JB087iB08p06668}

\bibitem[{{O'Keefe} \& {Ahrens}(1993)}]{OKeefe93}
---. 1993, \jgr, 98, 17011, \dodoi{10.1029/93JE01330}

\bibitem[{{Quintana} {et~al.}(2016){Quintana}, {Barclay}, {Borucki}, {Rowe}, \&
  {Chambers}}]{Quintana16}
{Quintana}, E.~V., {Barclay}, T., {Borucki}, W.~J., {Rowe}, J.~F., \&
  {Chambers}, J.~E. 2016, \apj, 821, 126, \dodoi{10.3847/0004-637X/821/2/126}

\bibitem[{{Reufer} {et~al.}(2012){Reufer}, {Meier}, {Benz}, \&
  {Wieler}}]{Reufer12}
{Reufer}, A., {Meier}, M. M.~M., {Benz}, W., \& {Wieler}, R. 2012, \icarus,
  221, 296, \dodoi{10.1016/j.icarus.2012.07.021}

\bibitem[{{Rizo} {et~al.}(2016){Rizo}, {Walker}, {Carlson}, {Horan},
  {Mukhopadhyay}, {Manthos}, {Francis}, \& {Jackson}}]{Rizo16}
{Rizo}, H., {Walker}, R.~J., {Carlson}, R.~W., {et~al.} 2016, Science, 352,
  809, \dodoi{10.1126/science.aad8563}

\bibitem[{{Rosenblatt} {et~al.}(2016){Rosenblatt}, {Charnoz}, {Dunseath},
  {Terao-Dunseath}, {Trinh}, {Hyodo}, {Genda}, \& {Toupin}}]{Rosenblatt16}
{Rosenblatt}, P., {Charnoz}, S., {Dunseath}, K.~M., {et~al.} 2016, Nature
  Geoscience, 9, 581, \dodoi{10.1038/ngeo2742}

\bibitem[{Rudge {et~al.}(2010)Rudge, Kleine, \& Bourdon}]{Rudge10}
Rudge, J.~F., Kleine, T., \& Bourdon, B. 2010, Nature Geoscience, 3, 439

\bibitem[{{Rufu} {et~al.}(2017){Rufu}, {Aharonson}, \& {Perets}}]{Rufu17}
{Rufu}, R., {Aharonson}, O., \& {Perets}, H.~B. 2017, Nature Geoscience, 10,
  89, \dodoi{10.1038/ngeo2866}

\bibitem[{{Stewart} \& {Leinhardt}(2012)}]{Stewart12}
{Stewart}, S.~T., \& {Leinhardt}, Z.~M. 2012, \apj, 751, 32,
  \dodoi{10.1088/0004-637X/751/1/32}

\bibitem[{{Tonks} \& {Melosh}(1993)}]{Tonks93}
{Tonks}, W.~B., \& {Melosh}, H.~J. 1993, \jgr, 98, 5319,
  \dodoi{10.1029/92JE02726}

\bibitem[{Williams \& Mukhopadhyay(2019)}]{Williams19}
Williams, C.~D., \& Mukhopadhyay, S. 2019, Nature, 565, 78,
  \dodoi{10.1038/s41586-018-0771-1}

\end{thebibliography}

\acknowledgments

We thank the anonymous reviewers for their constructive comments which have improved the quality of this manuscript. 
This work was supported by NASA grant 80NSSC18K0828 (PJC and STS). 
SJL gratefully acknowledges support from Harvard University's Earth and Planetary Sciences Department and Caltech's Division of Geological and Planetary Sciences. 
A summary of the simulations is available in the supplementary material. {The data used to produce the figures in this article, the EOS table for use with the modified version of GADGET-2, the input files for all the simulations discussed in this work, and code to read them are available from \url{https://doi.org/10.7910/DVN/YYNJSX} \citep{Carter19GIDATA}.} The modified GADGET-2 code is available in the online supplement of \citet{Cuk12}. The HERCULES code is available in the supplement of \citet{Lock17} and through the GitHub repositry:  \url{https://github.com/sjl499/HERCULESv1_user} \citep{HERCULES}. 

\clearpage
\appendix

\section{Modified specific impact energy}\label{a:qs}

The modified specific impact energy \citep{Lock17}, $Q_\mathrm{S}$, takes into account the efficiency with which energy is coupled into the impacting bodies, and is defined by,
\begin{equation}
    Q_\mathrm{S} = Q'_\mathrm{R} \left( 1 + {m \over M}\right) ( 1 - b ),
\end{equation}
where $Q'_\mathrm{R}$ is the specific impact energy in the centre of mass frame modified to include only the interacting mass of the projectile \citep[see][]{Leinhardt12}, $m$ and $M$ are the masses of the projectile and target, and $b$ is the impact parameter.

$Q'_\mathrm{R}$ is defined,
\begin{equation}
    Q'_\mathrm{R} = { \mu_\alpha \over \mu } Q_\mathrm{R},
\end{equation}
where the specific impact energy in the centre of mass frame, $Q_\mathrm{R}$, is given by, $Q_\mathrm{R} = 0.5 \mu v_\textrm{i}^2 / M_\mathrm{tot}$; $\mu$ is the reduced mass, $\mu = m M / M_\mathrm{tot}$; $v$ is the impact velocity in the centre of mass frame; and $M_\mathrm{tot}$ is the total mass. The modified reduced mass accounting only for the interacting portion of the projectile is given by,
\begin{equation}
    \mu_\alpha = { \alpha m M \over \alpha m + M},
\end{equation}
where $\alpha$ is the geometrically defined mass fraction of the projectile involved in the impact. $\alpha$ is given by,
\begin{equation}
    \alpha = {3 r l^2 - l^3 \over 4 r^3},
\end{equation}
where $l$ is the projected length of the projectile overlapping the target, and $r$ and $R$ are the radii of the projectile and target. $l$ takes one of two values, if $(R+r)b+r > R$ then $l = (1-b)(R+r)$, otherwise the whole projectile is involved and $l = 2r$ such that $\alpha=1$.

\clearpage


\clearpage

\end{document}